\documentstyle[aps,epsf,multicol]{revtex}
\begin{document}
\draft

\title{Application of the Bogolyubov's theory of weakly \\
non-ideal Bose gas on the A+A, A+B, B+B reaction-diffusion system}

\author{Zoran Konkoli}

\address{
  Department of Applied Physics \\
  Chalmers University of Technology \\ 
  and G\"oteborg University, \\
  SE 412 96 G\"oteborg, \\
  Sweden}

\date{\today}
\maketitle
\begin{abstract}
Theoretical methods for dealing with diffusion-controlled reactions
inevitably rely on some kind of approximation and to find the one that
works on a particular problem is not always easy. In here the
approximation used by Bogolyubov to study weakly non-ideal Bose gas,
to be refereed to as weakly non-ideal Bose gas approximation (WBGA),
is applied in the analysis of of the three reaction-diffusion models
(i) $A+A\longrightarrow \emptyset$, (ii) $A+B \longrightarrow
\emptyset$ and (iii) $A+A, B+B, A+B \longrightarrow \emptyset$ (the
ABBA model).  The two types of WBGA are considered, the simpler WBGA-I
and more complicated WBGA-II. All models are defined on the lattice to
facilitate comparison with computer experiment (simulation). It is
found that the WBGA describes A+B reaction well, it reproduces
correct $d/4$ density decay exponent. However, it fails in the case of
the A+A reaction and the ABBA model.  (To cure deficiency of WBGA in
dealing with A+A model the hybrid of WBGA and Kirkwood superposition
approximation is suggested.)  It is shown that the WBGA-I is identical
to the dressed tree calculation suggested by Lee in J. Phys. A {\bf
27}, 2633 (1994), and that the dressed tree calculation does not lead
to the d/2 density decay exponent when applied to the A+A reaction, as
normally believed, but it predicts d/4 decay exponent. Last, the usage
of the small $n_0$ approximation suggested by Mattis and Glasser in
Rev. Mod. Phys. {\bf 70}(3), 979 (1998) is questioned if used beyond
A+B reaction-diffusion model.
\end{abstract}

\begin{multicols}{2}
\narrowtext

\section{Introduction}

A variety of methods have been used to study diffusion-controlled
reactions (see, e.g.,
refs.~\cite{rev1,rev2,rev3,rev4,rev5,rev6,rev7,rev8} for review)
ranging from simplest pair-like or Smoluchowskii approach~\cite{OTB}
towards more sophisticated methods such as many particle density
function formalism~\cite{rev4,rev5} and field
theory~\cite{rev6,rev7,rev8}. In here, we focus on the field theory
approach which is exact but, like in any other theory, one needs
approximations to solve the problem. The particular way of making
approximate calculations will be analyzed, the Bogolyubov's theory of
weakly non-ideal Bose gas~\cite{bog}, to be referred to in the
following as Weakly non-ideal Bose Gas Approximation (WBGA).

The field theory is very attractive since it offers systematic way to
do calculation and there are well known procedures how to represent
stochastic dynamics of reaction-diffusion systems as field
theory. Such mapping can be carried out for lattice- and off-lattice
models, see \cite{rev6} and \cite{OTB} respectively for description.
Final result is that the dynamics is governed by a second quantized
Hamiltonian $H(a,a^\dagger)$,
\begin{equation}
  \frac{\partial}{\partial t} \Psi(t) = - H(a,a^\dagger) \Psi(t)
  \label{Hpsi}
\end{equation}
and system configuration at time $t$ can be extracted from state
vector $\Psi(t)$.  All observables can be expressed as a field
theoretic averages over products of creation and annihilation
operators $a^\dagger$ and $a$.

There is no need whatsoever to resort to the use of field theory as
all calculations could be done without it.  The reasons for using it
are mostly practical. Field theory is a powerful book-keeping
device. First, approximations involved can be made more transparent
and, second, due to this transparency it is relatively straightforward
to borrow approximations from other problems (and related forms of
field theory).

However, despite its elegance, when applied to the reaction-diffusion
systems, field theory seems to work with limited success. Most often
the only practical procedure of solving field theory is perturbative
calculation. Diffusion is set up as the zeroth order problem and
reaction is perturbation.  Problem is that in the most interesting
regime, when dimension of the system is low (bellow some critical
dimension), perturbation series diverges due to infra-red divergences.
Any non-field theoretic approach which aims at perturbative treatment
will suffer in a similar way.

The infra-read divergences are normally controlled in the framework of
the renormalization group (RG) technique.  When applying RG, the
initial density is relevant parameter and grows under renormalization
process.  Growing initial densities lead to infinities in expressions
which have to be controlled. So far, such control exists only in the
limited amount of cases, the A+A reaction being the simplest
example. Clearly, there is a need to avoid perturbative treatment
(together with RG) and look for alternative ways of calculation. In
here, focus is on the WBGA.

The WBGA was used to study reaction-diffusion systems
previously~\cite{rev8,OTB,ZO,BOP,GMY}.  In somewhat technical terms,
the basic idea behind WBGA is to neglect the products containing three
or higher number of annihilation operators having non-zero wave vector
in the Hamiltonian. This procedure leads to closed set of equations
for particle density and correlation functions. Solving these
equations amounts to resummation of infinite series of diagrams. In
this way WBGA appears as a non-perturbative technique which can be used
to control infra-red divergences of perturbation series. In the
following, we will distinguish between two types of approximations to
be refereed to as WBGA-I or WBGA-II which result in linear or
nonlinear set of equations of motion respectively.

On the more intuitive basis, the neglect of fluctuations on the
shorter length scale (large $k$ vectors) has to do with the slowness
of diffusion. For initial conditions when particle density is uniform
there is no $k\ne 0$ component in average particle density and the
density profile is flat. Since diffusion is a very slow process one
expects that this profile is kept all the time up to a small
perturbation. This picture is likely to be true when particle density
is low when compared to the reaction range, which is summarized in the
condition that $n r^d\ll 1$ where $n$ and $r$ denote the particle
density and the reaction range~\cite{OTB}. Naturally, there are other
scales in the problem (which change in time) and this kind of thinking
can not be employed without encountering problems~\cite{rev4}. In
reality, one really has to test the WBGA on the particular model to
see whether it works.

To analyze properties of the WBGA method, we apply it in the analysis
of three diffusion-controlled reactions (i) the A+A, (ii) the A+B and
(iii) the ABBA model. The ABBA model includes A+A, A+B and B+B
reactions simultaneously which allows for competitions between A+A and
A+B reactions (see ref.~\cite{KJ1,KJ2} for more details). The A+A and
A+B models have been studied by variety of methods (see, e.g.,
ref.~\cite{CGP} for review).  In here, the focus is on the WBGA
studies.  Apart from few initial studies of the A+A and A+B models the
WBGA have not been widely used. These early applications of WBGA on
A+A and A+B reaction-diffusion systems are mostly done for stationary
situation (and decay from it) and are listed in refs.~\cite{ZO} (A+A)
and \cite{BOP,GMY} (A+B). Also, the WBGA reappears in
paper~\cite{rev6} where it is used to study the A+B reaction.

The purpose of this work is twofold. First, we study A+A and A+B
reactions with uncorrelated initial conditions with particles
Poisson-distributed~\cite{foot2} at initial time $t=0$. In previous
work which employed WBGA method in refs.~\cite{ZO,BOP,GMY} it was
assumed that particles are initially correlated, with a initial
condition being prepared in a somewhat special way~\cite{foot1}.
Second, once the workings of the WBGA approximation is illuminated, we
apply the WBGA method to analyze the more complicated ABBA model. This
model was studied previously with field theoretic RG technique using
$\epsilon$-expansion and numerical simulation~\cite{KJ1,KJ2}. The ABBA
model describes a mixture of the A+A and A+B reactions which occur at
the same time. The A+A (A+B) component bias kinetics towards $d/2$
($d/4$) decay exponent and it is not a priori clear which exponent
will prevail. The analysis in~\cite{KJ1,KJ2} shows that both A and B
particles decay with $d/2$ decay exponent and different amplitudes.

The A+A model has been studied by variety of methods and lot is know
about it's behavior.  The study most relevant for this work is given
in ref.~\cite{Lee}. For uncorrelated initial conditions particle
density $n$ decays asymptotically as $n\sim {\cal A}(Dt)^{-d/2}$ for
$d<2$ while there is logarithmic correction at $d=2$ as $n\sim\ln(
Dt)/(8\pi Dt)$, where $d$ denotes dimension of system, $t$ is time and
$D$ denotes diffusion constant. Amplitude ${\cal A}$ is universal
number independent on the details of the model. Also, the exact
solution of the model is possible at $d=1$ where $n\sim 1/\sqrt{8\pi
Dt}$~\cite{Lush}.

The $d/2$ decay exponent of the A+A reaction is robust in a sense that
even simplest pair like approach (e.g. Smoluchowskii) predicts this
exponent. Thus, it is natural to test WBGA on the A+A model. It will
be shown that, somewhat surprisingly, WBGA fails to describe A+A
reaction: WBGA-I predicts $d/4$ decay exponent while WBGA-II gives
mean field exponent.

Taking slightly more complicated reaction scheme, such as the A+B,
leads to very hard problem where calculation methods start to differ
in their predictions. The nature of the asymptotics of A+B reaction
when initial number of A and B particles is equal has been hotly
debated for decades. For example, Smoluchowskii approach predicts $d/2$
decay exponent~\cite{rev4,rev5,OTB}. Field theoretic RG method predicts
correct $d/4$ decay for $3<d<4$~\cite{LeeCardy}, while for other
dimensions nothing can be said in the RG framework. Strangely enough,
seemingly the most sophisticated coupled-cluster technique argues for
mean field exponent~\cite{Rudav}. The true decay is governed by
$d/4$ exponent as shown by work of Bramson and Lebowitz which provided
strict mathematical proof~\cite{bram}. It will be shown that when
particles are initially uncorrelated, {\em i.e.}  Poisson-distributed,
both the WBGA-I and WBGA-II predict $d/4$ decay exponent. Strangely
enough, WBGA fails on simpler A+A reaction while it describes properly
more complicated A+B reaction.

The paper is organized as follows. In section \ref{sec:model} the
model to be studied is specified in detail, and mapping to the field
theory is described. The outcome of this section is Hamiltonian which
describes the most general two species reaction-diffusion model. In
section \ref{sec:EOM} equations of motion for density and correlation
functions are derived using WBGA method. In section \ref{sec:AA+WBGA}
the WBGA approach is used to describe A+A model, with application of
WBGA-I discussed in section \ref{sec:AA+WBGAI}, and application of
WBGA-II in section \ref{sec:AA+WBGAII}. Also in section
\ref{sec:AA+WBGAI} equivalence of dressed tree calculation with WBGA-I
is shown. Section \ref{sec:AA+hybrid} discusses modification of
Kirkwood superposition approximation in the spirit of WBGA. In section
\ref{sec:AB+WBGA} A+B model is studied by using WBGA-I and WBGA-II
methods. Finally, the analysis of the most complicated of three models
studied here, the ABBA model, is presented in section
\ref{sec:ABBA+WBGA}. Workings of WBGA-I are analyzed in section
\ref{sec:ABBA+WBGAI} and of WBGA-II in section
\ref{sec:ABBA+WBGAII}. The step by step merger of WBGA and Kirkwood
superposition approximation is discussed in \ref{sec:ABBA+hybrid}. The
summary and outline of future work is given in section
\ref{sec:conclusions}.

\section{The model and the mapping to the field theory}
\label{sec:model}

The mapping to the field theory will be carried out for the most
general two-species reaction-diffusion model.  The A+A, A+B and the
ABBA model will be obtained as special cases.  It will be assumed that
particles can not be created, neither by external source, nor by birth
process.  In the most general version of the two species
reaction-diffusion model each of the $A$ or $B$ particles jumps onto
the one of the neighboring lattice sites with rate (or diffusion
constant) $D_A$ or $D_B$, irrespectively of occupation number of the
site where particle jumps to.  Apart from diffusing particles
annihilate in pairs. Particle $\rho$ sitting at lattice site $x$ and
particle $\nu$ sitting at the site $y$ are assumed to annihilate with
rate $\sigma^{\rho\nu}_{xy}$. This is schematically denoted as
\begin{equation}
  \rho(x)+\nu(y)\rightarrow 0 
\end{equation}
where $\rho,\nu=A,B$ and $x,y=1,2,3,\ldots,V$. It is
assumed that there are $L$ lattice sites in one direction, and for $d$
dimensions total number of sites equals $V=L^d$. Labels $x$ and $y$
denote lattice sites.

The stochastic dynamics of the most general two-species model is
governed by the master equation of the system. The master equation
describes transitions between different configurations $c$. The
configuration of the system is specified by the occupancy of lattice
sites at certain time $t$;
\begin{equation}
  c\equiv(n_1,m_1,\ldots,n_x,m_x,\ldots,n_y,m_y,\ldots,n_V,m_V)
\end{equation}
$n_x$ and $m_x$ count number of A and B particles respectively at
$x$-th lattice site; $n_x,m_x=0,1,2,\ldots$ and $x=1,2,\ldots,V$. The
quantity of interest is $P(c;t)$, which denotes probability that at
given time $t$ system is in the configuration $c$. 

Diffusion and reaction contribute independently to change of $P(c;t)$;
\begin{equation}
  \frac{\partial}{\partial t}P(c;t) = 
     \dot P_{\rm D}(c;t) + \dot P_{\rm R}(c;t) 
  \label{ME}
\end{equation}
The influence of diffusion is described by
\begin{eqnarray}
  \dot P_{\rm D}(c;t) & = & 
      \sum_x\sum_{e(x)} \{ D_A [ (n_i+1) P(/n_x+1,n_e-1/;t) \nonumber \\
  & & - n_i P(c;t) ] + D_B [ (m_i+1) \nonumber \\
  & & \times P(/m_x+1,m_e-1/;t) - m_i P(c;t) ] \}
  \label{ME1}     
\end{eqnarray}
where notation $P(/\ldots/;t)$ indicates that content of the sites
specified by $/\ldots/$ has been modified in $c$. The $\sum_{e(x)}$
denotes sum over all first-neighbor sites of the site $x$. The change
of $P(c;t)$ driven by reactions is given by
\begin{eqnarray}
  & & \dot P_{\rm R}(c;t) = \sum_x \sigma^{AA}_{xx} \frac{(n_x+2)(n_x+1)}{2}
      P(/n_x+2/;t) \nonumber \\
  & & \ + \sum_x \sigma^{BB}_{xx} \frac{(m_x+2)(m_x+1)}{2}
      P(/m_x+2/;t) \nonumber \\
  & & \ + \sum_x \sigma^{AB}_{xx} (n_x+1)(m_x+1)
      P(/n_x+1,m_x+1/;t) \nonumber \\  
  & & \ + \sum_{x>y} \sigma^{AA}_{xy} (n_x+1)(n_y+1)
      P(/n_x+1,n_y+1/;t) \nonumber \\
  & & \ + \sum_{x>y} \sigma^{BB}_{xy} (m_x+1)(m_y+1)
      P(/m_x+1,m_y+1/;t) \nonumber \\
  & & \ + \sum_{x\ne y} \sigma^{AB}_{xy} (n_x+1)(m_y+1)
      P(/n_x+1,m_y+1/;t) \nonumber \\
  & & \ - \left[ 
            \sum_x\sigma^{AA}_{xx} \frac{n_x(n_x-1)}{2}
            + \sum_x\sigma^{BB}_{xx} \frac{m_x(m_x-1)}{2}  
           \right. \nonumber \\
  & & \ \ \ \ \ \ \ \  
            + \sum_x \sigma^{AB}_{xx} n_x m_x
            +  \sum_{x>y} \sigma^{AA}_{xy} n_x n_y 
            \nonumber \\
  & & \ \ \ \ \ \ \ \ 
            \left.
            + \sum_{x>y} \sigma^{BB}_{xy} m_x m_y 
            + \sum_{x\ne y} \sigma^{AB}_{xy} n_x m_y 
          \right] P(c;t) 
  \label{ME2}
\end{eqnarray}
The form of the master equation describing reactions can not be made
more compact since distinction has to be made between reaction on the
same and different lattice sites. For example, the term describing A+A
reaction at the same lattice site has the pre-factor $n_x(n_x-1)$ while
for two different sites another form $n_x n_y$ has to be used.

The master equation (\ref{ME}) is the first order differential
equation and one has to specify initial condition in the form $P(c;0)$
for all $c$. In the following it will be assumed that $P(c;0)$ is
Poisson distribution with averages denoted by $n_{0,A}$ and
$n_{0,B}$ for $A$ and $B$ particles respectively. This means that
$P(c;0)$ factorises,
\begin{equation}
  P(c;0) = \Pi_{x=0,1,2,\ldots,V} p(n_x,n_{0,A}) p(m_x,n_{0,B}) 
\end{equation}
where $p(n,\bar n)$ denotes Poisson distribution function
\begin{equation}
  p(n,\bar n) = e^{-\bar n} \bar n^n/n! \ , \ \ n=0,1,2,\ldots
\end{equation}

Eq.~(\ref{ME}) can be translated into a Schr\"odinger-type equation
(\ref{Hpsi}) by defining state vector,
\begin{eqnarray}
 &  \Psi(t) = & \sum_{n_1,m_1,...,n_V,m_V}  
                P(n_1,m_1,...,n_V,m_V;t) \nonumber \\ 
 &           & \ \ \ \ \ \times (a_1^\dagger)^{n_1} (b_1^\dagger)^{m_1} ... 
               (a_V^\dagger)^{n_V} (b_V^\dagger)^{m_V} 
            |0\rangle 
  \label{Psi}
\end{eqnarray}
where $a_x^\dagger$, $a_x$, $b_x^\dagger$, and $b_x$ are creation and
annihilation operators for A and B particles respectively;
\begin{eqnarray}
  & & [a_x,a^\dagger_y] = \delta_{x,y} \, , \ 
      [b_x,b^\dagger_y] = \delta_{x,y} \, , \nonumber \\  
  & & [a_x,b^\dagger_y] = [a^\dagger_x,b_y] = 0 
\end{eqnarray}
and $[{\rm I},{\rm II}]\equiv {\rm I}\ {\rm II} - {\rm II}\ {\rm I}$
denotes commutator.  One can see that Eq.~(\ref{ME}) is equivalent to
the Eq.~(\ref{Hpsi}) provided dynamics is governed by the
second-quantized Hamiltonian $H = H_{\rm D} + H_{\rm R}$. Term
$H_{\rm D}$ describes diffusion
\begin{equation}
  H_{\rm D} = \sum_x \sum_{e(x)} 
                 [ D_A a^\dagger_x ( a_x - a_e ) + 
                   D_B b^\dagger_x ( b_x - b_e ) ]
  \label{Hdiffx}
\end{equation}
and $H_{\rm R}$ reactions
\begin{eqnarray}
  H_{\rm R} & = & 
        \frac{1}{2} \sum_{x,y} \sigma^{AA}_{xy}
          (a^\dagger_x a^\dagger_y-1) a_x a_y + \nonumber \\
 & &  + \frac{1}{2} \sum_{x,y} \sigma^{BB}_{xy}
          (b^\dagger_x b^\dagger_y-1) b_x b_y + \nonumber \\
 & &  + \sum_{x,y} \sigma^{AB}_{xy}(a^\dagger_x b^\dagger_y-1) a_x b_y
  \label{Hreactx}
\end{eqnarray}
Please note that expression for $H_{\rm R}$ has the more compact
form than the expression for related part of the master equation
(\ref{ME2}) as all sums in (\ref{Hreactx}) over $x$ and $y$ are
unrestricted (a first sign of convenient book-keeping). Also the fact
that reactions A+A, B+B and A+B happen independently from each other
influences the form of $H_{\rm R}$; contributions describing A+A
(first line of Eq.~\ref{Hreactx}), B+B (second line) and A+B (third
line) enter additively.

Once $H$ has been specified $\Psi(t)$ can be found from (\ref{Hpsi}),
\begin{equation}
  \Psi(t) = e^{-H t} \Psi_0
  \label{PsiHPsi0}
\end{equation}
In general, form of $\Psi_0$ depends on the initial particle
distribution and for Poisson-distributed particles equals
\begin{equation}
  \Psi_0 = {\rm exp} 
           \left[ 
              n_{0,A}\sum_x ( a_x^\dagger - 1) + 
              n_{0,B}\sum_x ( b_x^\dagger - 1 ) 
           \right] 
           |0\rangle
\end{equation}
Since $\Psi(t)$ is available (at least in principle) one can calculate
various observables with following recipe
\begin{equation}
  \langle O(n_x(t)) \rangle = 
  \langle 1 | O(a^\dagger_x a_x) |\Psi(t)\rangle
  \label{obs}
\end{equation}
where $O(n_x)$ denotes any function which depends on particle numbers
in arbitrary way. The generalization of Eq.~(\ref{obs}) for more
complicated form for $O(n_x,m_y,\ldots)$ is trivial. Vector $\langle
1|$ is given by
\begin{equation}
  \langle 1| \equiv \langle 0| {\rm exp} \left[ \sum_x ( a_x + b_x ) \right] 
  \label{bra1}
\end{equation}
being left eigenvector of $a^\dagger_x$ and $b^\dagger_x$ with
eigenvalue $1$.  To see that Eq.~(\ref{obs}) indeed evaluates
observables correctly one can expand function $O$ in Taylor series in
$n_x$ and check that for every term in Taylor series
Eq.~(\ref{obs}) works.

Following ref.~\cite{rev6}, since $\langle 1 |a^\dagger_x \ne 0$, it is
useful to transform Eq.~(\ref{obs}) slightly in order to obtain form
where $\langle 1 |$ changes into the vacuum state $\langle 0 |$. This
makes calculation somewhat easier since one can use property of the
vacuum that $\langle 0 | a^\dagger_x = 0$. To do the transformation
following identity proves useful,
\begin{equation}
  \langle 1 | O(a^\dagger_x, a_x ) |0\rangle = 
  \langle 0 | O(a^\dagger_x+1, a_x ) |0\rangle  
  \label{obs1}
\end{equation}
and by using (\ref{obs1}) Eq.~(\ref{obs}) can be written in the form
\begin{equation}
  \langle O(n_x(t)) \rangle = 
  \langle 0 | O((a^\dagger_x+1)a_x) e^{-\bar Ht}|\bar\Psi_0\rangle
  \label{obs2}
\end{equation} 
where bar over $H$ and $\Psi_0$ indicates that substitution
\begin{equation}
  a^\dagger_x\rightarrow a^\dagger_x+1 \ , \
  b^\dagger_x\rightarrow b^\dagger_x+1 \ , \ \ x=0,1,2,...,V
  \label{adag1}
\end{equation}
have been made. The $\bar\Psi_0$ is given by
\begin{equation}
  \bar\Psi_0 = {\rm exp} 
           \left[ 
              n_{0,A}\sum_x a_x^\dagger  + 
              n_{0,B}\sum_x b_x^\dagger  
           \right] 
           |0\rangle
\end{equation}
and $\bar H=\bar H_{\rm D}+\bar H_{\rm R}$. After the shift
$H_{\rm D}$ does not change the form, i.e. $H_{\rm D}=\bar
H_{\rm D}$, while $H_{\rm R}$ changes into
\begin{eqnarray}
  \bar H_{\rm R} & = & 
        \frac{1}{2} \sum_{x,y} \sigma^{AA}_{xy}
          (a^\dagger_x + a^\dagger_y + a^\dagger_x a^\dagger_y) 
          a_x a_y + \nonumber \\
 & &  + \frac{1}{2} \sum_{x,y} \sigma^{BB}_{xy}
          (b^\dagger_x + b^\dagger_y + b^\dagger_x b^\dagger_y) 
          b_x b_y + \nonumber \\
 & &  + \sum_{x,y} \sigma^{AB}_{xy}
        (a^\dagger_x + b^\dagger_y + a^\dagger_x b^\dagger_y) a_x b_y
  \label{Hreactx'}
\end{eqnarray}
In the following we compactify notation and use
\begin{equation}
  \langle f(a_x,a^\dagger_y,\ldots) \rangle \equiv
  \langle 0 | f(a_x,a^\dagger_y,\ldots) 
  e^{-\bar H t}| \bar\Psi_0\rangle 
\end{equation}
where $f$ denotes any function of operators $a_x$, $a^\dagger_y$,
$b_x$ and $b^\dagger_y$ for any $x$ and $y$. Also notation 
\begin{equation}
  \bar\Psi(t) \equiv e^{-\bar H t} \bar\Psi_0
  \label{BarPsiHPsi0}
\end{equation}
will be useful.

To avoid effects of boundaries periodic boundary conditions are
assumed and one can introduce Fourier transforms for operators,
\begin{eqnarray}
 & &  a_x = \frac{1}{\sqrt{V}} \sum_k e^{ikx} a_k \, , \  
      a^\dagger_x = \frac{1}{\sqrt{V}} \sum_k e^{-ikx} a^\dagger_k 
      \label{akx}\\
 & &  b_x = \frac{1}{\sqrt{V}} \sum_k e^{ikx} b_k \, , \ 
      b^\dagger_x = \frac{1}{\sqrt{V}} \sum_k e^{-ikx} b^\dagger_k
      \label{bkx} 
\end{eqnarray}
and reaction rates
\begin{equation}
  \sigma^{\rho\nu}_x = \frac{1}{V} \sum_k \sigma^{\rho\nu}_k e^{ikx} \, , \ 
    \sigma^{\rho\nu}_k = \sum_x \sigma^{\rho\nu}_x e^{-ikx} 
  \label{sigmakx} 
\end{equation}
with $\rho,\nu=A,B$. For convenience, Fourier transforms are defined
slightly differently for operators and reaction rates (just to reduce
explicit occurrence of $V$ in expressions later on). Also, it is useful
to express Hamiltonian $\bar H$ in terms of creation and annihilation
operators in $k$-space. Starting from (\ref{Hdiffx}) and
(\ref{Hreactx'}), and using (\ref{akx})-(\ref{sigmakx}) gives
\begin{equation}
  H_{\rm D} =   D_A \sum_k k^2 a^\dagger_k a_k 
                 + D_B \sum_k k^2 b^\dagger_k b_k + {\cal O}(k^4) 
  \label{Hdiffk}
\end{equation}
and
\begin{eqnarray}
 \bar H_{\rm R} & = & 
        \frac{1}{\sqrt{V}} 
        \sum_{q,k} \sigma^{AA}_{q} a^\dagger_k a_{k-q} a_q \nonumber \\
 & &  + \frac{1}{2V} \sum_{q,k,l} \sigma^{AA}_q 
          a^\dagger_k a^\dagger_l a_{k-q} a_{l+q} \nonumber \\
 & &  + \frac{1}{\sqrt{V}} \sum_{q,k} \sigma^{BB}_{q} b^\dagger_k 
            b_{k-q} b_q \nonumber \\
 & &  + \frac{1}{2V} \sum_{q,k,l} \sigma^{BB}_q 
            b^\dagger_k b^\dagger_l b_{k-q} b_{l+q}  \nonumber \\
 & &  + \frac{1}{\sqrt{V}} \sum_{q,k} \sigma^{AB}_{q} 
           ( a^\dagger_k a_{k-q} b_q + b^\dagger_k b_{k-q} a_q )  
           \nonumber \\
 & &  + \frac{1}{V} \sum_{q,k,l} \sigma^{AB}_q 
            a^\dagger_k b^\dagger_l a_{k-q} b_{l+q}
  \label{Hreactk}
\end{eqnarray}
Please note that Eq.~(\ref{Hdiffk}) is an approximation which is valid
for small $k$.  The form in Eq.~(\ref{Hreactk}) is exact. Once the
explicit form of $\bar H$ is known one can proceed with calculation of
particle density, as shown in the next section.

\section{Equations of motion for density and the correlation functions}
\label{sec:EOM}

Equation~(\ref{obs2}) summarizes how observables are calculated within
the field theory formalism.  We continue with the specific case of
local particle densities $n_A(x,t)$ and $n_B(x,t)$, which can be
calculated with $O=a^\dagger_x a_x$ and $O=b^\dagger_x b_x$
respectively; using (\ref{obs2}) one gets $n_A(x,t)=\langle a_x
\rangle$ and $n_B(x,t)=\langle b_x \rangle$, or more explicitly
\begin{eqnarray}
 & &  n_A(x,t) = \langle 0 | a_x | \bar\Psi \rangle = 
                 \langle 0 | a_x e^{-\bar H t} | \bar\Psi_0 \rangle  
      \label{nax} \\ 
 & &  n_B(x,t) = \langle 0 | b_x | \bar\Psi \rangle = 
                 \langle 0 | b_x e^{-\bar H t} | \bar\Psi_0 \rangle  
      \label{nbx}  
\end{eqnarray}

If initial conditions are translationally invariant, which is the case
for Poisson-distributed particles, local particles densities are
position independent, i.e. $n_A(x,t)=n_A(t)$ and
$n_B(x,t)=n_B(t)$. Also at $t=0$ $n_\rho(0)=n_{0,\rho}$ with
$\rho=A,B$. It is convenient to re-express $n_A(t)$ and $n_B(t)$ as
$n_A(t)\equiv\frac{1}{V}\sum_x n_A(x,t)$ and
$n_B(t)\equiv\frac{1}{V}\sum_x n_B(x,t)$. Using field theory $n_A(t)$
and $n_B(t)$ can be calculated from
\begin{eqnarray}
  & & n_A(t) \equiv \frac{1}{V} \langle \sum_x a_x \rangle 
             =  \frac{1}{\sqrt{V}} \langle a_0 \rangle 
             \label{nAkx} \\
  & & n_B(t) \equiv \frac{1}{V} \langle \sum_x b_x \rangle
             =  \frac{1}{\sqrt{V}} \langle  b_0 \rangle 
             \label{nBkx}
\end{eqnarray}
where the sum over $x$ divided by $\sqrt{V}$ was recognized as $k=0$
component of $a_k$ and $b_k$.

The equations of motion for $n_A(t)$ and $n_B(t)$ can be derived as
follows. $n_A(t)$ and $n_B(t)$ are proportional to $\langle a_0
\rangle$ and $\langle b_0\rangle$ respectively, and it is sufficient
to derive equations for them. 
Taking time derivative of $\langle a_0 \rangle$
and $\langle b_0 \rangle$ gives,
\begin{eqnarray}
  & & \frac{\partial}{\partial t} \langle a_0 \rangle 
      =  -\langle [a_0,\bar H] \rangle \\
  & & \frac{\partial}{\partial t} \langle b_0 \rangle 
      =  -\langle [b_0,\bar H] \rangle
\end{eqnarray}
The commutator emerges when time derivative acts on ${\rm exp}(-\bar H
t)$~\cite{foot3}. Evaluating the commutator with $\bar H$ given in
(\ref{Hdiffk}) and (\ref{Hreactk}), and taking into account
(\ref{nAkx}) and (\ref{nBkx}) gives
\begin{eqnarray}
  \frac{\partial n_A}{\partial t} & = & 
     - \left[ 
          \sigma^{AA}_0 n_A n_A + \sigma^{AB}_0 n_A n_B + 
       \right. \nonumber \\ 
    & & \ \ \ 
           \left. 
                + \frac{1}{V} \sum_{k\ne 0} 
                ( \sigma^{AA}_k \Gamma_k^{AA} + \sigma^{AB}_k \Gamma_k^{AB})
           \right] 
       \label{dnAdt} \\
  \frac{\partial n_B}{\partial t} & = & 
     - \left[ 
          \sigma^{BB}_0 n_B n_B + \sigma^{AB}_0 n_A n_B + 
       \right. \nonumber \\ 
    & & \ \ \ 
            \left. 
                + \frac{1}{V} \sum_{k\ne 0} 
                ( \sigma^{BB}_k \Gamma_k^{BB} + \sigma^{AB}_k \Gamma_k^{AB})
            \right] 
       \label{dnBdt}
\end{eqnarray}
where 
\begin{eqnarray}
  \Gamma_k^{AA} & \equiv & \langle a_k a_{-k}\rangle  \\ 
  \Gamma_k^{BB} & \equiv & \langle b_k b_{-k}\rangle  \\
  \Gamma_k^{AB} & \equiv & \langle a_k b_{-k}\rangle
\end{eqnarray}
To derive (\ref{dnAdt}) and (\ref{dnBdt}) the sum over $k$ is split
into $k=0$ and $k\ne 0$ parts and assumption is made that $k=0$
components are non-fluctuating, {\em i.e.} $\langle a_0 a_0 \rangle
\approx \langle a_0 \rangle \langle a_0 \rangle$ and likewise for
$\langle a_0 b_0 \rangle$ and $\langle b_0 b_0 \rangle$ (thermodynamic
limit).

The equations (\ref{dnAdt}) and (\ref{dnBdt}) involve correlation
functions $\Gamma_k^{\rho\nu}$ with $\rho,\nu=A,B$ which we proceed to
calculate. The time evolution of correlators is governed by $[O,\bar H]$
with $O=a_ka_{-k},\ a_k b_{-k},\ b_k b_{-k}$. Evaluating commutators
gives equations of motion,
\begin{eqnarray}
  & &  \frac{\partial}{\partial t} \Gamma_k^{AA} =
       - 2 D_A k^2 \Gamma_k^{AA}   
       - \left[  
         \sigma^{AA}_k n_A^2 + \right. \nonumber \\
  & &  \ \ \ \ 
         + \frac{1}{V} \sum_{q\ne k} 
         \sigma^{AA}_q \Gamma^{AA}_{k-q}   
         + 2(\sigma^{AA}_0+\sigma^{AA}_k) n_A \Gamma^{AA}_k + \nonumber \\
  & &  \ \ \ \ + \left. 
         \sigma^{AB}_k n_A ( \Gamma^{AB}_{k} + \Gamma^{AB}_{-k} )
         + 2 \sigma^{AB}_0 n_B \Gamma^{AA}_k  \right]
       \label{dtGa} \\
  & &  \frac{\partial}{\partial t} \Gamma_k^{BB} =
       - 2 D_B k^2 \Gamma_k^{BB}   
       - \left[  
         \sigma^{BB}_k n_B^2 + \right. \nonumber \\
  & &  \ \ \ \ 
         + \frac{1}{V} \sum_{q\ne k} 
         \sigma^{BB}_q \Gamma^{BB}_{k-q}   
         + 2(\sigma^{BB}_0+\sigma^{BB}_k) n_B \Gamma^{BB}_k + \nonumber \\
  & &  \ \ \ \ + \left. 
         \sigma^{AB}_k n_B ( \Gamma^{AB}_{k} + \Gamma^{AB}_{-k} )
         + 2 \sigma^{AB}_0 n_A \Gamma^{BB}_k  \right]        
       \label{dtGb} \\ 
  & &  \frac{\partial}{\partial t} \Gamma_k^{AB} =
       - ( D_A + D_B ) k^2 \Gamma_k^{AB} 
       - \left[ \sigma^{AB}_k n_A n_B + \right. \nonumber \\
  & &  \ \ \ \ 
         + \left. \frac{1}{V} \sum_{q\ne k} 
         \sigma^{AB}_q \Gamma^{AB}_{k-q} \right]   
         - \left[  
             ( \sigma^{AA}_0 + \sigma^{AA}_k ) n_A + \right. \nonumber \\
  & &  \ \ \ \ 
            \left.  + ( \sigma^{BB}_0 + \sigma^{BB}_k ) n_B 
           + \sigma^{AB}_0 ( n_A + n_B ) \right] \Gamma^{AB}_k - \nonumber \\
  & &  \ \ \ \ 
       - \sigma^{AB}_k ( n_A \Gamma^{BB}_k + n_B \Gamma^{AA}_k ) 
       \label{dtGc}
\end{eqnarray}
Eqs.~(\ref{dtGa}), (\ref{dtGb}) and (\ref{dtGc}) are approximate since
correlators of type $\langle a_k a_q a_l \rangle$ with $k\ne 0$, $q\ne
0$, and $l\ne 0$ are assumed to be small. This is exactly the content of 
the WBGA.

Few comments about the inversion symmetry (in $k$-space) of
$\Gamma^{\rho\nu}_k$ $\rho,\mu=A,B$ are in order. By construction,
$\Gamma^{\rho\rho}_k(t) = \Gamma^{\rho\rho}_{-k}(t)$ for $\rho=A,B$
since these functions represent correlations for same operator
types. It is not obvious whether this property holds for
$\Gamma^{AB}_k$ which describes correlations for different operators
type. Using Eq.~(\ref{dtGc}) to derive equation of motion for
$\Gamma^{(-)}_k(t)\equiv \Gamma^{AB}_k(t) - \Gamma^{AB}_{-k}(t)$ one
can see that $\Gamma^{(-)}_k(t)=0$ is solution provided
$\Gamma^{(-)}_k(0)=0$. The initial conditions for $\Gamma^{\rho\nu}_k$
with $\rho\nu=A,B$ are given by
\begin{equation}
  \Gamma^{\rho\nu}_k(0)=\delta_{k,0} V n_{0,\rho} n_{0,\nu} \ , \ \ 
                        \rho,\nu=A,B
\end{equation}
where $\delta_{x,y}$ denotes Kronecker delta-function,
\begin{equation}
  \delta_{x,y} =
    \left\{
      \begin{array}{ll}
        1 & x=y \\ [5pt]
        0 & x\ne y
      \end{array}
    \right.
  \label{krondeltaxy}
\end{equation}
Also, notation
\begin{equation}
  \bar\delta_{x,y}=1-\delta_{x,y}
\end{equation} 
will be used.  Thus $\Gamma^{(-)}_k(0)$ is indeed zero. It follows
that $\Gamma^{AB}_k(t)=\Gamma^{AB}_{-k}(t)$ for every $t\ge 0$. In the
following whenever $\Gamma^{AB}_k + \Gamma^{AB}_{-k}$ appears in the
equations of motion, we will use assumption of inversion symmetry and
shorten expression to $2\Gamma^{AB}_k$. Naturally, if this inversion
symmetry is broken at $t=0$ one has to keep full form
$\Gamma^{AB}_k+\Gamma^{AB}_{-k}$ in the equations of motion, but such
case is not considered here.

The last terms in Eqs.~(\ref{dtGa})-(\ref{dtGc}), which are product of
density and correlator, and appear to be third order in density ${\cal
O}(n^3)$, lead to non-linear equations of motion. These terms come
from averages of the type $\langle a_k a_q a_l\rangle$ where one of
$\{k,q,l\}$ momenta is zero while remaining two are not.  One example
would be $\langle a_0 a_k a_l\rangle$ which is approximatively equal
to $\langle a_0 \rangle \langle a_k a_l \rangle$.  (Same discussion
would apply irrespectively which operator one uses in average, $a_k$
or $b_k$.) Being third order in density, it is tempting to neglect
these ${\cal O}(n^3)$ terms, since one expects that leading
contribution should come from terms which are second order in density
${\cal O}(n^2)$, e.g. first and second terms on the right hand side of
Eq.~(\ref{dtGa}) and likewise for Eqs.~(\ref{dtGb}) and
(\ref{dtGc}). To test the effect of neglecting or keeping ${\cal
O}(n^3)$ terms two approximations will be studied WBGA-I or WBGA-II
with ${\cal O}(n^3)$ terms taken away or kept in the calculation.  WBGA-I
approximation results in a linear set of equations which makes
the analytical analysis possible. Also, as formulated above, the WBGA-II
and WBGA appear to be equivalent. In the following, the term WBGA will
imply both WBGA-I and WBGA-II.

The equations of motion for the three models we wish to study, the
A+A, A+B and ABBA are easily extracted from the most general form
given in (\ref{dnAdt})-(\ref{dnBdt}) and (\ref{dtGa})-(\ref{dtGc}). To
obtain equations of motion for the ABBA model one simply sets
$D_A=D_B=D$ and
\begin{equation}
  \sigma^{AA}_{xy}=\sigma^{BB}_{xy}=\lambda\delta_{x,y} \ , \ \ 
  \sigma^{AB}_{xy}=\delta\delta_{x,y}
  \label{sigmas}
\end{equation} 
An obvious distinction is being made between Kronecker delta-function
$\delta_{x,y}$ and the symbol $\delta$ which denotes the reaction
rate. Thus particles have to meet at the same lattice site in order to
react.  Further, to obtain the A+A model one simply sets $\delta=0$
(this decouples A+A and B+B reactions, i.e. particles A and B move and
react independently of each other). To get A+B model one takes
$\lambda=0$ which rules out the A+A reaction. In the following section
we continue with analysis of the A+A model within WBGA framework.

\section{WBGA applied to the A+A reaction}
\label{sec:AA+WBGA}

Using (\ref{sigmas}) with $\delta=0$ in (\ref{dnAdt})-(\ref{dnBdt})
and (\ref{dtGa})-(\ref{dtGc}) gives equation of motion for the density
\begin{equation}
  \frac{\partial n}{\partial t} = - \lambda [ n^2 + \Phi ] 
  \label{dn1}
\end{equation}
and the correlator
\begin{equation}
  \frac{\partial}{\partial t} \Gamma_k = 
       - 2 D k^2 \Gamma_k 
       - \lambda [ n^2 + \Phi]
       - 4 \lambda n \Gamma_k 
  \label{dtG1} 
\end{equation}
Letter $A$ has been dropped on $n_A$ and $\Gamma_k^A$ to simplify
notation, and likewise $n_{0,A}$ is shortened to $n_0$. $\Phi(t)$ is
implicitly defined by correlators,
\begin{equation} 
  \Phi(t)=\frac{1}{V}\sum_{k\ne 0}\Gamma_k(t)
  \label{Phi}
\end{equation}
Thus equations above are meant to describe the model where only one
type of species, A, jumps on the lattice and particles have a chance
to react only when at the same lattice site. The equations above will
be solved in the following two subsections using WBGA-I and WBGA-II
approaches.

\subsection{The WBGA-I approximation}
\label{sec:AA+WBGAI}

In the WBGA-I approximation, when term proportional to $n \Gamma_k$ is
dropped, Eq.~(\ref{dtG1}) can be studied analytically.  Uncorrelated
(Poisson-like) initial condition is described by
$\Gamma_k=\delta_{k,0}Vn_0^2$ and solution of (\ref{dtG1}), with $k\ne
0$, reads
\begin{equation}
  \Gamma_k(t) = - \lambda \int_0^t dt' e^{-2Dk^2(t-t')}[n(t')^2 + \Phi(t')]
  \label{G1k}
\end{equation}
Please note that the $\Gamma_0(t)$ is determined from
$\Gamma_0(t)=Vn(t)^2$ (thermodynamic limit) and not from (\ref{G1k}).
Summation of Eq.~(\ref{G1k}) over $k\ne 0$ and division by $V$
gives
\begin{equation}
  \Phi(t) = -\lambda \int_0^t dt' G(t-t') [ n(t')^2 + \Phi(t') ]
  \label{PhiIntPhi}
\end{equation}
where
\begin{equation}
  G(t-t') \equiv \frac{1}{V}\sum_{k\ne 0}e^{-2k^2D(t-t')}
  \label{Gtt}
\end{equation}
was introduced.  As shown in the appendix \ref{app:Gtt}, for large
lattice size when $V\rightarrow\infty$, expression above can be
approximated as
\begin{equation}
  G(t-t') \approx 
    \left[
       8\pi D (t-t' + \eta )
    \right]^{-d/2}
  \label{Gtt3}
\end{equation}
where $\eta = \frac{1}{8\pi D}$.

Equations (\ref{dn1}), (\ref{PhiIntPhi}) and (\ref{Gtt3}) completely
specify $n(t)$. It is not possible to solve them analytically,
however, large time behavior of $n(t)$ can be extracted. To do this
we introduce $\varphi\equiv n^2+\Phi$ and rewrite Eqs.~(\ref{dn1}) and
(\ref{PhiIntPhi}) as
\begin{eqnarray}
  & & \frac{\partial n}{\partial t} = - \lambda \varphi \label{dn3} \\
  & & \varphi(t) = n(t)^2 - \lambda 
      \int_0^t dt' G(t-t') \varphi(t') \label{varphi}
\end{eqnarray}
which completely specify $n(t)$.

By using Laplace transform it is possible to transform equations
(\ref{dn3}) and (\ref{varphi}) into a single equation.  Laplace
transform is defined as
\begin{equation}
  X(s)=\int_0^\infty dt e^{-st}X(t)
\end{equation}
For $X=n,\varphi$ same symbol will be used for Laplace transform as
for the original function. The only exception to the rule are two
cases. For $X(t)=n(t)^2$, $X(s)=n_2(s)$, while for $X(t)=G(t)$,
$X(s)=g(s)$. 

Taking Laplace transform of Eq.~(\ref{varphi}) one gets $\varphi(s) =
n_2(s) - \lambda g(s) \varphi(s)$, and combining it with $\varphi(s) =
(sn(s)-n_0)/\lambda$ from (\ref{dn3}) gives
\begin{equation}
  n_2(s) = \left[ g(s)+\frac{1}{\lambda} \right] [ n_0 - s n(s) ]
  \label{ns1}
\end{equation}
The $g(s)$ is the Laplace transform of $G(t)$,
\begin{equation}
  g(s) = (8\pi D)^{-d/2} e^{\eta s} s^{d/2-1} \Gamma(1-d/2,\eta s)
\end{equation}
$\Gamma(\beta,x)$ denotes incomplete Gamma function;
\begin{equation}
  \Gamma(\beta,x) = \int_x^\infty du\, u^{-1+\beta} e^{-u}
\end{equation}
The analytic continuation of $\Gamma(\beta,x)$ is possible. For
non-integer $\beta$ and $\beta=0$, $\Gamma(\beta,z)$ is
multiple-valued function of $z$ with a branch point at $z=0$, and has
no poles. Dividing by $g(s) + 1/\lambda$ Eq.~(\ref{ns1}) results in
\begin{equation}
  s n(s) - n_0 = - \lambda_{\rm eff}(s) n_2(s) 
  \label{ns}
\end{equation}
where $\lambda_{\rm eff}(s)$ denotes Laplace transform of the
effective reaction rate,
\begin{equation}
  \lambda_{\rm eff}(s) = \frac{\lambda}{1+\lambda g(s)}
  \label{lambda_s}
\end{equation}
Finally, taking inverse Laplace transform of (\ref{ns}) gives
\begin{equation}
  \frac{\partial n}{\partial t} = - \int_0^t dt' 
     \lambda_{\rm eff}(t-t') n(t')^2
  \label{dtc}
\end{equation}

Equations (\ref{dn3})-(\ref{varphi}) and Eq.~(\ref{dtc}) are fully
equivalent.  Both equations have been dealt with before, however, in
a very different context. Eq.~(\ref{dtc}) was obtained in
ref.~\cite{Lee} through a diagrammatic technique and referred to as the
{\em dressed-tree} calculation. Thus in here we have shown that WBGA-I
is equivalent to the dressed tree
calculation. Eqs.~(\ref{dn3})-(\ref{varphi}) have been also obtained in
the study of the entirely different A+B model in ref.~\cite{rev6}.

Studies \cite{Lee} and \cite{rev6} suggest contradictory result:
ref.~\cite{Lee} argues that the dressed tree calculation gives $d/2$
decay exponent for particle density, while ref.~\cite{rev6} argues for
the $d/4$ decay exponent.  In ref.~\cite{Lee} it was incorrectly
concluded that dressed-tree calculation results in the $d/2$ exponent,
which basically came from balancing wrong terms in Laplace transformed
version of Eq.~(\ref{dtc}). In here calculation done in
ref.~\cite{Lee} will be repeated to show how to balance terms
correctly. Also, calculation will justify approximations employed in
ref.~\cite{rev6} more rigorously as the method of calculation employs
Laplace Transform and well know Tauberian theorems which relate small
$s$ with large $t$ behavior. In this way all the approximations are
controlled.

To extract asymptotic behavior for $n(t)$ from Eq.~(\ref{dtc}) one
assumes that at large times density decays as
\begin{equation}
  n(t) \approx {\cal A} (\mu + t)^{-\alpha}
  \label{nmu}
\end{equation}
$\cal A$ and $\alpha$ denote amplitude and exponent of decay to be
found.  $\mu$ is introduced as regulator for small $t$ so that Laplace
transform of $n(t)$ and $n^2(t)$ exist;
\begin{eqnarray}
  & & n(s) = {\cal A} e^{s\mu} s^{\alpha-1} \Gamma(1-\alpha,\mu s) 
      \label{n1s} \\
  & & n_2(s) = {\cal A}^2 e^{s\mu} s^{2\alpha-1} \Gamma(1-2\alpha,\mu s)
      \label{n2s}
\end{eqnarray}
and please note that $n(s)^2\ne n_2(s)$. To extract asymptotics one
inserts (\ref{n1s})-(\ref{n2s}) and (\ref{lambda_s}) into (\ref{ns}) and
expands in small $s$ (to extract leading order behavior for large
$t$) and matches the most dominant terms.

The expansion of $g(s)$ for small $s$ is given by 
\begin{eqnarray}
 g(s)  & = & (8\pi D)^{-d/2} e^{s\eta} \times \nonumber \\
 & & \times \left[ \Gamma(1-d/2) s^{d/2-1} + 
            \frac{2\eta^{1-d/2}}{d-2} + {\cal O}(s) \right] 
 \label{gsa}
\end{eqnarray}
for $d\ne 2,4,6,\ldots$.  For $d=2$ one has
\begin{eqnarray}
 g(s)  & = & (8\pi D)^{-1} e^{s\eta} 
             \left[ -\gamma_E - \ln ( \eta s ) + {\cal O}(\eta s) \right] 
 \label{gsb}
\end{eqnarray}
where $\gamma_E$ is Euler constant.  Please note that the behavior of
$g(s)$ for small $s$ is qualitatively different for $d<2$ and $d>2$
which has to do with recurrence of random walks bellow and above
$d=2$. For small $s$ and $d<2$ $g(s)\propto s^{d/2-1}$ while for $d>2$
$g(s)=const$. At $d=2$ there is logarithmic dependence on $s$. The
term $e^{s\eta}$ can be neglected if leading order behavior for small
s (large $t$) is sought for.

At the moment we focus on the $d<2$ case.  Inserting approximate
formulas above for $g(s)$ into (\ref{lambda_s}) gives
\begin{equation}
  \lambda_{\rm eff}(s) \sim 
          \frac{(8\pi D)^{d/2}}{\Gamma(1-d/2)} s^{-d/2+1} \ , \ \ d<2 
  \label{lambda_s1}
\end{equation}
Since value for $\alpha$ is not known one has to separate various
cases: expansion for $n(s)$ reads
\begin{equation}
 n(s) = {\cal A} \left\{ 
    \begin{array}{ll}
      \left[ \Gamma(1-\alpha) s^{\alpha-1} 
             + {\cal O}(1) \right] & \alpha<1 \\ [5pt]
       \left[ \frac{\mu^{1-\alpha}}{\alpha-1} 
             + {\cal O}(s^{\alpha-1}) \right] & \alpha > 1
    \end{array}
    \right.
  \label{n1s_app}
\end{equation}
and likewise for $n_2(s)$
\begin{equation}
 n_2(s) = {\cal A}^2 \left\{ 
    \begin{array}{ll}
      \left[ \Gamma(1-2\alpha) s^{2\alpha-1} 
             + {\cal O}(1) \right] & 2\alpha<1 \\ [5pt]
      \left[ \frac{\mu^{1-2\alpha}}{2\alpha-1} 
             + {\cal O}(s^{2\alpha-1}) \right] & 2\alpha>1
    \end{array}
    \right.
  \label{n2s_app}
\end{equation}
Inserting small $s$ expansions (\ref{lambda_s1})-(\ref{n2s_app}) into
(\ref{ns}) gives
\begin{eqnarray}
  & & {\cal A} [s^\alpha \Gamma(1-\alpha) + {\cal O}(s)] - n_0 =  \nonumber \\
  & & - {\cal A}^2 \frac{(8\pi D)^{d/2}}{\Gamma(1-d/2)}  
      \left[ s^{2\alpha-d/2} \Gamma(1-2\alpha) + {\cal O}(s^{1-d/2}) \right] 
  \label{match}
\end{eqnarray}
Also, please note that there are two different forms to use for $n(s)$
and $n_2(s)$ in (\ref{n1s_app}) and (\ref{n2s_app}) and the ones used
in (\ref{match}) were for $\alpha<1$ and $2\alpha<1$ respectively
(same choice was made in ref.~\cite{Lee}). Once $\alpha$ is found, one
has to check these conditions on $\alpha$ for self consistency.  There
are two ways to match the terms in (\ref{match}), (a) as in the
ref.~\cite{Lee}, and (b) in a way related to the work in
ref.~\cite{rev6} . We begin with first case.

Balancing $s^\alpha$ term on the left hand side of (\ref{match}) with
$s^{2\alpha-d/2}$ on the right hand side, gives $\alpha=d/2$ and 
\begin{equation}
  {\cal A}_a=-\frac{1}{\pi}{\rm sin}(\pi d)
              \Gamma(d)\Gamma(1-d/2)^2(8\pi D)^{-d/2}
\end{equation}
Also from $\alpha<1$ and $2\alpha<1$ one has constraint that
$d<1$. However, for $d<1$ the term ${\rm sin}(\pi d)$ is positive
which makes amplitude ${\cal A}_a$ negative. Thus all physical
conditions can not be met with this type of matching. In
ref.~\cite{Lee} the condition $d<1$ [coming from the fact that first
row is used in (\ref{n2s_app})] was overlooked (if $d>2$ is allowed
amplitude ${\cal A}_a$ is perfectly acceptable).

The $\alpha=d/2$ scenario can still turn out to be true. With this
choice of $\alpha$ and $d<2$ condition coming from (\ref{lambda_s1})
the second row in (\ref{n2s_app}) have to be used. Again, carrying out
similar type of matching procedure would give negative
amplitude. Finally, the $\alpha=d/2$ avenue has to be given up.

At this stage one is left by the second (b) way of balancing, {\em i.e.}
matching constant $n_0$ term on the left hand side of (\ref{match})
with $s^{2\alpha-d/2}$ on the right hand side. (The remaining terms,
{\em e.g.} such as the $s^\alpha$ on the left hand side can be
balanced by considering sub-leading corrections to $n(s)$). This way of
balancing immediately gives $\alpha=d/4$ and
\begin{equation}
  {\cal A}_b=\sqrt{n_0}(8\pi D)^{-d/4}
  \label{Ab}
\end{equation}
with constraints that $d<2$ (Eq.~\ref{lambda_s1} was used
to get Eq.~\ref{match}).

Matching the constant term $n_0$ on the left hand side of
(\ref{match}) is rather counter-intuitive since in the framework of
Laplace transform constant can normally be disregarded when large $t$
behavior is sought for. To see how this comes about it is useful to
turn back to Eq.~(\ref{ns}).

Equation~(\ref{ns}) comes from Eq.~(\ref{ns1}). For simplicity reasons
we focus on the case $\lambda=\infty$ in (\ref{ns1}). It is clear
that at the right hand side of equation (\ref{ns1}) the $s n(s)$ term
is sub-leading to $n_0$. (True enough, $n_0$ is constant but it is
multiplied by $g(s)$.) Thus $n_2(s)$ indeed has to be matched with
$g(s) n_0$. This procedure results in amplitude ${\cal A}_b$ obtained
previously. Also, by using form (\ref{ns1}), one can show that
amplitude ${\cal A}_b$ as given in Eq.~(\ref{Ab}) is valid even for
$d>2$.  Analysis can be repeated with finite value of $\lambda$ with
the same outcome.  Before proceeding, it is important to mentioned
that procedure outlined above does not work at $d\ge 4$ and it has to
be modified.

The main findings of this subsection are twofold. First, it has been
shown that dressed tree calculation is equivalent to WBGA-I. Second,
it was shown that WBGA-I (and dressed tree calculation) fail to
describe A+A reaction predicting $d/4$ decay exponent;
\begin{equation}
  n(t) \sim \sqrt{n_0} (8\pi Dt)^{-d/4}
  \label{ntasym}
\end{equation}

All findings of this subsection are summarized in Fig.~1. The
numerical treatment of Eqs.~(\ref{dn3}) and (\ref{varphi}) confirms
asymptotic decay given in Eq.~(\ref{ntasym}). Equations (\ref{dn3})
and (\ref{varphi}) were solved previously numerically in
ref.~\cite{rev6}. In here, the features of the decay curves are
somewhat different from the ones obtained in ref.~\cite{rev6}. For
example, curves shown in this work have concave form (bend upward),
while curves in fig. 1 of ref.~\cite{rev6} are convex (bend downward)
as if asymptotics have not yet been reached. Also, in here, there is
no intersection of curves, which can be found in
ref.~\cite{rev6}. These difference could come from the numerical
treatment. Apart from using double precision, to obtain curves in
Fig. 1 the method of integration was used which integrates exactly
$\int_0^t dt'(t-t')^{-d/2}f(t')$ provided $f(t)$ is piecewise linear
in $t$. The more detailed description of numerical treatment is shown
in the appendix \ref{app:kernal}

In the next subsection it will be shown that, in the case of A+A
reaction, weaknesses of WBGA-I method extend to WBGA-II level.

\subsection{The WBGA-II approximation}
\label{sec:AA+WBGAII}

When term $n\Gamma_k$ is kept in Eq.~(\ref{dtG1}), equivalent of
Eq.~(\ref{varphi}) reads
\begin{equation}
  \varphi(t) = n(t)^2 - \lambda \int_0^t  dt' I(t,t') \varphi(t')
  \label{varphi1}
\end{equation}
while Eq.~(\ref{dn3}) stays the same. The $I(t,t')$ is given by
\begin{equation}
  I(t,t') = G(t-t')\ {\rm exp} 
     \left[ -4\lambda \int_{t'}^t dt'' n(t'') \right] 
  \label{Itt}
\end{equation}
The asymptotics of Eq.~(\ref{varphi1}) can not be extracted by Laplace
transform, and it is more convenient to use approach of
ref.~\cite{rev6}. For large $t$, Eq.~(\ref{varphi1}) can be
approximated by
\begin{equation}
  \varphi(t) \approx n(t)^2 - I(t,0) {\cal I}(t)
  \label{varphiapprox}
\end{equation}
where ${\cal I}(t)\equiv \lambda \int_0^t dt' \varphi(t')$.  This step
is valid provided two conditions are satisfied. First, the term
$I(t,t')$ has to vanish as time difference $t-t'$ grows. Second, the
integral $\int_0^\infty dt \varphi(t)$ has to be finite. Using
(\ref{dn3}) one gets ${\cal I}(t)=n_0 - n(t) \approx n_0$ and
(\ref{varphiapprox}) becomes
\begin{equation}
  \frac{\partial n}{\partial t} \approx -\lambda n^2 + \lambda I(t,0) n_0
  \label{dndtapprox}
\end{equation}
Equation above is solved with the assumption that asymptotically
$n(t)\sim {\cal A}/t$, which is checked self consistently at the
end. Using postulated asymptotics for $t$, one can see from
Eq.~(\ref{Itt}) that $I(t,0)\sim {\rm const}\ t^{-(d/2+4\lambda{\cal
A})}$. Assuming that
\begin{equation}
  \frac{I(t,0)}{n(t)^2}\rightarrow 0  \ , \ \ t\rightarrow\infty
  \label{Iasym}
\end{equation}
one can solve (\ref{dndtapprox}) in form $\partial n/\partial
t=-\lambda n^2$, and get ${\cal A}=1/\lambda$. Assumption
(\ref{Iasym}) is correct provided $2< d/2+4\lambda{\cal A}=d/2+4$,
which is true for any $d$. This shows that $n(t)\approx 1/(\lambda t)$
is asymptotic form for the solution of Eqs.~(\ref{dn3}) and
(\ref{varphi1}). This means that the last term ($n\Gamma_k$) in
(\ref{dtG1}) only influences intermediate behavior when $t$ is not too
large. For large $t$, WBGA-II gives exactly the same asymptotics as the
pure mean field treatment.

Thus main finding so far is that both WBGA-I and WBGA-II fail to
describe A+A reaction.  This is somewhat surprising as even simplest
pair-approach, e.g. Smoluchowskii method, describes exponent of A+A
correctly.  Clearly A+A reaction can not be viewed as a weakly
interacting Bose-gas. The question is what is the minimum modification
of WBGA which will provide correct result for the A+A model? This
question will be answered in the next subsection.

\subsection{The hybrid of WBGA and Kirkwood superposition approximation}
\label{sec:AA+hybrid}

To see how to improve WBGA one has to clarify what went wrong in the
first place.  We start from problematic equation (\ref{dtGa}) which
becomes (\ref{dtG1}) when terms with $\sigma^{AB}_k$ are set to
zero. To trace why WBGA fails it is useful to rewrite Eq.~(\ref{dtGa})
as it looks one step before WBGA is made, and we keep only terms
describing A+A reaction:
\begin{eqnarray}
  \frac{\partial}{\partial t} \Gamma_k & = &
       - 2 D k^2 \Gamma_k   
       - \left[  
           \sigma_k n^2 
           + \frac{1}{V} \sum_{q\ne k} \sigma_q \Gamma_{k-q} 
         \right] - \dot\Gamma^{(3)}
   \label{dtGa1}
\end{eqnarray}
where
\begin{equation}
   \dot\Gamma^{(3)} = \frac{1}{\sqrt{V}}
          \sum_q \sigma_q ( \langle a_{-k} a_{k-q} a_q \rangle 
                          + \langle a_{k} a_{-k-q} a_q \rangle )
   \label{dotGam}
\end{equation}
is focus of present subsection.  In technical terms, the usage of
WBGA can be translated into approximating three point density $\langle
a_x a_y a_z \rangle$ in a particular way. In the case of WBGA-I one
simply takes
\begin{equation}
  \langle a_{k_1} a_{k_2} a_{k_3} \rangle=
  \langle a_x a_y a_z \rangle= 0
  \label{a3i}
\end{equation}
while in WBGA-II one assumes
\begin{eqnarray}
  \langle a_{k_1} a_{k_2} a_{k_3} \rangle & \approx &  
       \delta_{k_1,0} \delta_{k_2,0} \delta_{k_3,0} a_0^3 + 
       \nonumber \\
   & & + \delta_{k_1,0} \bar\delta_{k_2,0} \bar\delta_{k_3,0}
         a_0 \langle a_{k_2} a_{k_3} \rangle +   
       \nonumber \\
   & & + \bar\delta_{k_1,0} \delta_{k_2,0} \bar\delta_{k_3,0} 
         a_0 \langle a_{k_1} a_{k_3} \rangle +
        \nonumber \\
   & & + \bar\delta_{k_1,0} \bar\delta_{k_2,0} \delta_{k_3,0} 
         a_0 \langle a_{k_1} a_{k_2} \rangle
   \label{a3iik} 
\end{eqnarray}
where $\bar\delta_{k,0}\equiv 1-\delta_{k,0}$. Inserting (\ref{a3iik})
into (\ref{dtGa1}) gives the terms describing A+A process in
(\ref{dtGa}). It is useful to transform approximation above into the
x-space to understand nature of approximation better. Inverse Fourier
transform of (\ref{a3iik}) gives~\cite{foot4}
\begin{eqnarray}
  \langle a_x a_y a_z \rangle & \approx &  
    n^3 + n ( \langle a_x a_y \rangle - n^2 ) + \nonumber \\
    & & + n ( \langle a_x a_z \rangle - n^2 ) 
        + n ( \langle a_y a_z \rangle - n^2 )
  \label{a3iix} 
\end{eqnarray}
The WBGA approximations given in Eqs.~(\ref{a3i}) and (\ref{a3iix})
can be contrasted to the Kirkwood superposition approximation,
\begin{equation}
  \langle a_x a_y a_z \rangle \approx \frac{1}{n^3} 
    \langle a_x a_y \rangle \langle a_x a_z \rangle 
    \langle a_y a_z \rangle
    \label{kirkwood}
\end{equation}
which is known to describe A+A problem correctly (at least the decay
exponent). 

It is interesting to note that instead of (\ref{kirkwood}) one could
carry out Kirkwood superposition approximation in a different way,
e.g. as $\langle n_x n_y n_z \rangle \approx \langle n_x n_y \rangle
\langle n_x n_z \rangle \langle n_y n_z\rangle/n^3$. (There is no such
ambiguity when only single occupancy of site is allowed.)  
When multiple site occupancy is allowed, as considered here, we argue
that Eq.~(\ref{kirkwood}) is more accurate way to implement Kirkwood
superposition approximation.

By looking at equation (\ref{a3iix}) it is possible to understand WBGA
better. Eq.~(\ref{a3iix}) suggests that WBGA is somewhat equivalent to
the additive expansion of correlation functions. It has been argued
that such additive approximation is inferior to the Kirkwood
superposition approximation~\cite{rev4,rev5}. This in turn explains
why WBGA fails to describe the A+A reaction.

Clearly, to correctly describe A+A reaction one has to use Kirkwood
superposition approximation, which we modify further in the spirit of
WBGA, assuming that higher order products of annihilation operators
$a_k,b_k$ with $k\ne 0$ are small, and accounting for thermodynamic
limit (extracting $a_0$ or $b_0$ out of field theoretic
averages). This is done in two stages. First, Kirkwood superposition
approximation in Eq.~(\ref{kirkwood}) is rephrased in the k-space
which gives~\cite{foot5}
\begin{equation}
  \langle a_{k_1} a_{k_2} a_{k_3} \rangle \approx 
     \frac{\delta(k_1+k_2+k_3)}{n^3V^{3/2}} 
     \sum_l \Gamma_l \Gamma_{k_1-l} \Gamma_{k_2+l}
  \label{kirkwoodk}
\end{equation}
To get the improved form for three body terms one inserts
(\ref{kirkwoodk}) into (\ref{dotGam}) which gives
\begin{equation}
  \dot\Gamma^{(3)} \approx \frac{2}{n^3V^2} 
     \sum_{q,l} \sigma_q \Gamma_l \Gamma_{k+l} \Gamma_{k-q+l} 
  \label{dotGam1}
\end{equation}
The expression above was obtained by using symmetry properties of
$\Gamma_k=\Gamma_{-k}$ and $\sigma_q=\sigma_{-q}$. Also upon inserting
(\ref{kirkwoodk}) into (\ref{dotGam}) the two terms on the right hand side
of (\ref{dotGam}) contribute equally resulting in factor two in
(\ref{dotGam1}).  Equation above is too complicated to be used in
practise and we approximate it further. 

In the spirit of WBGA the terms in (\ref{kirkwoodk}) which contain
large number of correlation functions with $k$-vector different from
zero are neglected. This is done in two stages, first sum over $q$ is
split into $q=k$ and $q\ne k$ parts, and then for each of sums various
contributions from sum over $l$ are distilled to extract
non-fluctuating $k=0$ operators. This gives
\begin{eqnarray}
  & & \dot\Gamma^{(3)} \approx 
        \frac{2}{n^3V^2} \left[  \sigma_k ( \Gamma_0^2 \Gamma_k 
     + \Gamma_0 \Gamma_k^2) + \right. \nonumber \\
  & & \left. + \sum_{q\ne k} \sigma_q \Gamma_0 ( \Gamma_k \Gamma_q  
       + \Gamma_k \Gamma_{k-q} + \Gamma_q \Gamma_{q-k} ) \right] 
  \label{dotGam2} 
\end{eqnarray}
where terms of the type $\Gamma_k\Gamma_q\Gamma_l$ with $k,q,l\ne 0$
have been neglected.  Eq.~(\ref{dotGam2}) is still complicated to be
used, at least to obtain analytic result. First term on the right hand
side of (\ref{dotGam2}) can be neglected since it leads to mean field
behavior (as shown in section \ref{sec:AA+WBGAII}). Second term can be
absorbed into the one of the three terms under the sum sign,
.e.g. under the first term. Second and third terms under the sum sign
couple correlation functions in a non-trivial way and are neglected in
the following for simplicity reasons. With $\Gamma_0\approx n^2V$, and
applying recipe just described gives
\begin{equation}
  \dot\Gamma^{(3)} \approx \frac{2}{nV} \Gamma_k \sum_{q} \sigma_q \Gamma_q  
  \label{dotGam3} 
\end{equation}
which is midway between WBGA-II (Eq.~\ref{a3iik}) and Kirkwood
superposition approximation (Eq.~\ref{kirkwoodk}). It is interesting
to note that equation above is very close to the shortened Kirkwood
superposition approximation discussed in refs.~\cite{rev4,rev5}. Using
approximation above to decouple three body density, and particular
form for $\sigma^{AA}_{xy}$ used throughout this section, gives following
equations of motion,
\begin{equation}
  \frac{\partial}{\partial t} \Gamma_k = 
       - 2 D k^2 \Gamma_k 
       - \lambda ( n^2 + \Phi )
       - 2 \lambda \Gamma_k \frac{ \Phi + n^2 }{n} 
  \label{dtG2} 
\end{equation}
which should be contrasted with Eq.~(\ref{dtG1}). The most convenient
way to solve (\ref{dn1}) and (\ref{dtG2}) is to introduce $\chi_k$ as
$\Gamma_k=n^2\chi_k$ and $n^2+\Phi = n^2\chi$ where
\begin{equation}
  \chi\equiv 1+\frac{1}{V}\sum_{k\ne 0}\chi_k
\end{equation}
$\chi_k$ with $k=0$ is set equal to $V$ and does not change in time
(thermodynamic limit). Applying change of variables just described
modifies Eq.~(\ref{dn1}) into
\begin{equation}
  \frac{\partial}{\partial t} n(t) = - \kappa(t) n(t)^2  
  \label{dn2} 
\end{equation}
where effective reaction rate 
\begin{equation}
  \kappa(t)=\lambda\chi(t)
\end{equation}
was introduced. Same change of variables transforms Eq.~(\ref{dtG2})
into
\begin{equation}
  \frac{\partial}{\partial t} \chi_k = 
       - 2 D k^2 \chi_k - \lambda \chi
  \label{chi}
\end{equation}
Equation (\ref{chi}) can be solved for all $\chi_k$ and $k\ne 0$
(pretending that $\chi(t)$ is known), and after summing over $k\ne 0$
one gets following integral equation
\begin{equation}
  \chi(t) = 1 - \lambda \int_0^t dt' G(t-t') \chi(t')
\end{equation}
Solution of the equation above can be found by Laplace transform which gives
\begin{equation}
  \kappa(s) = \frac{\lambda}{s[1+\lambda g(s)]}
  \label{kappachis}
\end{equation}
Also, one can integrate Eq.~(\ref{dn2}) which gives
\begin{equation}
  n(t) = \frac{n_0}{1+ n_0 \bar\kappa(t) } \ , \ \ 
  \bar\kappa(t) = \int_0^t dt' \kappa(t') 
  \label{nt}
\end{equation} 
It is not possible to obtain closed expression for $\kappa(t)$ and
$n(t)$.  However, asymptotic form of $n(t)$ can be extracted.

Inserting small $s$ expansion of $g(s)$ (see Eqs.~\ref{gsa} and
\ref{gsb}) into (\ref{kappachis}) gives
\begin{equation}
  \kappa(s) \sim 
     \left\{ 
       \begin{array}{ll}
          \frac{(8\pi D)^{d/2}}{\Gamma(1-d/2)} s^{-d/2} & d<2 \\ [5pt]
          - \frac{8\pi D}{s[\gamma_E+\ln(\eta s)]} & d=2 \\ [5pt]
          \frac{\lambda}{1+\lambda g(0) }s^{-1} & d>2
       \end{array}
     \right.
  \label{kappas}
\end{equation}
for $s\rightarrow 0$. Taking inverse Laplace transform of equation
above gives leading order behavior for effective reaction rate constant,
\begin{equation}
  \kappa(t)\sim 
    \left\{ 
      \begin{array}{ll}
         \frac{(8\pi D)^{d/2}}{\Gamma(1-d/2)\Gamma(d/2)} 
             t^{d/2-1} & d<2 \\ [5pt]
          \frac{8\pi D}{\ln t/\eta} & d=2 \\ [5pt]
         \frac{\lambda}{1+\lambda g(0)} & d>2 
      \end{array}
    \right.
  \label{kappat}
\end{equation}
when $t\rightarrow\infty$.  To find inverse Laplace transform of
$\kappa(s)$ for $d=2$ (second line Eq.~\ref{kappas}) is somewhat
involved, please see appendix \ref{app:kappa} for details.  Finally,
inserting (\ref{kappat}) into (\ref{nt}) gives following asymptotics
\begin{equation}
  n(t) \sim \left\{ \begin{array}{ll}
              \Gamma(1-d/2) \Gamma(1+d/2) (8\pi Dt)^{-d/2} & d<2 \\ [5pt]
              \frac{\ln 8\pi Dt}{8\pi Dt} & d = 2 \\ [5pt]
              \left[ \frac{1}{\lambda} + g(0) \right] t^{-1} & d>2
            \end{array} \right.
\end{equation}
where $g(0)$ entering in the third row can easily be found from
(\ref{gsa}).

\section{WBGA applied to the A+B reaction}
\label{sec:AB+WBGA}

Equations of motion for density and correlation functions describing
A+B model result from (\ref{dnAdt})-(\ref{dnBdt}) and
(\ref{dtGa})-(\ref{dtGc}) by using (\ref{sigmas}) with
$\lambda=0$. For simplicity reasons, we focus on $n_A$=$n_B\equiv n$
case and omit labels $A$ and $B$. Also, as in the previous section
$n_0=n_{0,A}=n_{0,B}$. Applying procedure outlined above leads to the
equations for particle densities
\begin{equation}
  \frac{\partial n}{\partial t} = - \delta ( n^2 + \Phi_c ) 
  \label{dn4}
\end{equation}
where $\Phi_c$ is given by the AB correlation function,
\begin{equation}
  \Phi_c(t) = \frac{1}{V} \sum_{k\ne 0}\Gamma_k^{AB}(t)
\end{equation}
Equations for correlators $\Gamma_k\equiv\Gamma_k^{AA}=\Gamma_k^{BB}$
and $\Gamma_k^c\equiv\Gamma_k^{AB}$ are given by
\begin{eqnarray}
  & & \left( \frac{\partial}{\partial t} + 2Dk^2 \right) \Gamma_k^c =
     - \delta ( n^2 + \Phi_c ) 
      -2 \delta n ( \Gamma_k + \Gamma_k^c ) \\
  & & \left( \frac{\partial}{\partial t} + 2Dk^2 \right) \Gamma_k = 
      -2 \delta n ( \Gamma_k + \Gamma_k^c ) 
\end{eqnarray}
The most convenient way to solve equations above is to diagonalise
them by subtraction and addition. The final result is that
correlations of AB pairs are governed by
\begin{equation}
  \Phi_c(t) = - \delta \int_0^t dt' [ G(t,t') + I(t,t') ] 
                 [n(t')^2+\Phi_c(t')]
  \label{corAB}
\end{equation}
and for AA (or BB) pairs as
\begin{equation}
  \Phi(t) = \delta \int_0^t dt' [ G(t,t') - I(t,t') ] [n(t')^2+\Phi_c(t')]
  \label{corAABB}
\end{equation}
The $I(t,t')$ appearing in equations (\ref{corAB}) and (\ref{corAABB})
has the same form as in (\ref{Itt}) with trivial change of $\lambda$
into $\delta$. Same type of analysis as in the subsection
\ref{sec:AA+WBGAII} leads to conclusion that approximation
$G(t,t')+I(t,t')\approx G(t,t')$ can be used, which in turn leads to
$d/4$ density decay exponent.  Interestingly enough, both WBGA-I and
WBGA-II approaches lead to correct $d/4$ exponent when used to solve
A+B model.

In some sense the WBGA approach seems to be suited rather well for A+B
reaction. Quite the contrary can be said for A+A reaction as shown
previously. To understand workings of WBGA on the more general model
the ABBA model will be studied in the next section.

\section{WBGA applied to the ABBA model}
\label{sec:ABBA+WBGA}

The ABBA model was suggested in refs.~\cite{KJ1,KJ2} and it has a
number of interesting properties. The model is obtained as the special
case of the general two species model discussed in section
\ref{sec:model} where A+A and B+B reactions occur with same reaction
rate $\lambda$, while A+B with reactions rate $\delta$, and
$\delta>\lambda$ is taken. The fact that A and B species have equal
diffusion constants is very important since it leads to survival of
minority species. Minority species is the one with smaller
concentration at $t=0$, and we chose B to be minority,
i.e. $n_{0,A}>n_{0,B}$. By survival it is meant that particle density
ratio saturates to constant $u(t)\equiv n_A(t)/n_B(t)\rightarrow{\rm
const}$ as $t\rightarrow\infty$. (The mean field analysis predicts
vanishing of minority species $u(t)\rightarrow\infty$ as
$t\rightarrow\infty$.  Also, similar model has been studied in
ref.~\cite{Howard} where it was shown that if $D_A\ne D_B$ only one
type of species survives.) 

The goal of the present section is to understand strengths and
weaknesses of WBGA approach by testing it on the ABBA model which
combines A+A and A+B reactions.  These reactions have been studied in
previous sections but now they are allowed to occur simultaneously. It
is interesting to see how WBGA approach performs in such situation.
Clearly, there are plenty of possibilities how to combine A+A and A+B
reactions, but the way they are combined in the ABBA model gives rise
to many interesting effects. (For example, there is survival of
minority species, recovery of initial density ratio, dependence of
amplitudes on initial densities and reaction rates~\cite{foot6},
please see refs.~\cite{KJ1,KJ2} for details).

We start from equations of motion given in section \ref{sec:EOM} which
describe very general two species reaction-diffusion model.
Assumptions in Eq.~(\ref{sigmas}) describe content of the ABBA
model. Using (\ref{sigmas}) in Eqs.~(\ref{dnAdt})-(\ref{dnBdt}) gives
\begin{eqnarray}
  & & \frac{\partial}{\partial t} n_A = 
        - \left( 
             \lambda n_A^2 + \delta n_A n_B + 
             \lambda \Phi_{AA} + \delta \Phi_{AB} 
          \right) 
          \label{dtnA} \\
  & & \frac{\partial}{\partial t} n_B = 
        - \left( 
             \lambda n_B^2 + \delta n_A n_B + 
             \lambda \Phi_{BB} + \delta \Phi_{AB} 
          \right)
          \label{dtnB} 
\end{eqnarray}
where
\begin{equation}
  \Phi_{\rho\nu} = \frac{1}{V} \sum_{k\ne 0} \Gamma^{\rho\nu}_k \ , \ \ 
  \rho,\nu=A,B
\end{equation}
Also, Eqs. for correlators (\ref{dtGa})-(\ref{dtGc}) simplify to
\begin{eqnarray}
  \frac{\partial}{\partial t} \Gamma_k^{AA} 
      & = & - 2 D k^2 \Gamma_k^{AA}   
            - \lambda \left(  n_A^2 + \Phi_{AA} \right) - \nonumber \\
      &   & - 2 ( 2 \lambda n_A + \delta n_B ) \Gamma^{AA}_k 
            - 2 \delta n_A \Gamma^{AB}_k 
      \label{GammaAA} \\
  \frac{\partial}{\partial t} \Gamma_k^{BB} 
      & = & - 2 D k^2 \Gamma_k^{BB}   
            - \lambda \left(  n_B^2 + \Phi_{BB} \right) - \nonumber \\
      &   & - 2 ( 2 \lambda n_B + \delta n_A ) \Gamma^{BB}_k 
            - 2 \delta n_B \Gamma^{AB}_k  
      \label{GammaBB} \\
  \frac{\partial}{\partial t} \Gamma_k^{AB} 
      & = & - 2 D k^2 \Gamma_k^{AB}   
            - \delta \left(  n_A n_B + \Phi_{AB} \right) - \nonumber \\
      &   & - ( 2 \lambda + \delta ) (n_A+n_B) \Gamma^{AB}_k - \nonumber \\
      &   & - \delta ( n_A \Gamma^{BB}_k + n_B \Gamma^{AA}_k )
      \label{GammaAB}
\end{eqnarray}
The equations above will be solved in the next two subsections within
WBGA-I and WBGA-II approaches.

\subsection{The WBGA-I approximation}
\label{sec:ABBA+WBGAI}

In the framework of WBGA-I all seemingly ${\cal O}(n^3)$ terms of the
type $n_\rho \Gamma^\nu_k$ with $\rho=A,B$ and $\nu=AA,AB,BB$ are
thrown away in (\ref{GammaAA})-(\ref{GammaAB}). Following similar
steps as in subsection \ref{sec:AA+WBGAI} gives
\begin{eqnarray}
 & & \frac{\partial}{\partial t} n_A 
       = - ( \lambda \varphi_{AA} + \delta \varphi_{AB} ) 
     \label{dadt}\\
 & & \frac{\partial}{\partial t} n_B 
       = - ( \lambda \varphi_{BB} + \delta \varphi_{AB} )
     \label{dbdt}
\end{eqnarray}
and
\begin{eqnarray}
  & & \varphi_{AA} = n_A^2  - \lambda \int_0^t dt' G(t-t') \varphi_{AA}(t') 
      \label{dvaraa} \\
  & & \varphi_{BB} = n_B^2 - \lambda \int_0^t dt' G(t-t') \varphi_{BB}(t') 
      \label{dvarbb} \\
  & & \varphi_{AB} = n_A n_B - \delta \int_0^t dt' G(t-t') \varphi_{AB}(t')
      \label{dvarab}
\end{eqnarray}
where $\varphi_{\rho\nu}\equiv n_\rho n_\nu + \Phi_{\rho\nu}$ for
$\rho,\nu \in \{A,B\}$.  To solve Eqs.~(\ref{dadt})-(\ref{dvarbb}) it
is possible to employs same technique as in section
\ref{sec:AA+WBGAII}. Equations above can be approximated by
\begin{eqnarray}
  & & 0 \approx n_A^2 - G(t,0) {\cal I}_{AA}(t)
      \label{zeroaa} \\
  & & 0 \approx n_B^2 - G(t,0) {\cal I}_{BB}(t)
      \label{zerobb} \\
  & & 0 \approx n_A n_B - G(t,0) {\cal I}_{AB}(t)
      \label{zeroab}
\end{eqnarray}
where
\begin{equation}
{\cal I}_{\rho\nu}(t) \equiv  
    ( \lambda\delta_{\rho,\nu} + \delta \bar\delta_{\rho,\nu})
    \int_0^t dt' \varphi_{\rho\nu}(t')\
    , \ \ \rho=A,B 
\end{equation}
By integrating Eqs.~(\ref{dadt}) and (\ref{dbdt}) a useful
relationship can be derived for ${\cal I}_{\rho\nu}$, $\rho,\nu=A,B$;
\begin{eqnarray}
  & & {\cal I}_{AA}(t) + {\cal I}_{AB}(t) = n_{0,A}-n_A(t) \approx n_{0,A}
      \label{IAA} \\ 
  & & {\cal I}_{BB}(t) + {\cal I}_{AB}(t) = n_{0,B}-n_B(t) \approx n_{0,B}
      \label{IBB}
\end{eqnarray}
Using (\ref{zeroaa})-(\ref{zeroab}) gives $( n_A + n_B )^2 \approx
G(t) ({\cal I}_{AA}+{\cal I}_{BB}+2{\cal I}_{AB})$, and adding
(\ref{IAA}) and (\ref{IBB}) finally gives 
\begin{equation}
  (n_A + n_B)^2 \approx G(t) ( n_{0,A}+n_{0,B} )
  \label{nab+}
\end{equation}
Also from (\ref{zeroaa}) and (\ref{zerobb})
$n_A^2 - n_B^2 \approx G(t) ( {\cal I}_{AA}-{\cal I}_{BB})$ and
subtraction of (\ref{IAA}) and (\ref{IBB}) gives 
\begin{equation}
  n_A^2 - n_B^2 \approx G(t) ( n_{0,A}-n_{0,B} )
  \label{nab-}
\end{equation}
After solving (\ref{nab+}) and (\ref{nab-}) for $n_A$ and $n_B$ one
gets
\begin{eqnarray}
  n_\rho \sim \frac{n_\rho(0)}{\sqrt{n_{0,A}+n_{0,B}}} (8\pi Dt)^{-d/4} 
  \ , \ \ \rho=A,B
\end{eqnarray}

According to WBGA-I both particles decay with $d/4$ exponent and
amplitudes given above.  The WBGA-I predicts same decay exponent
as for the pure A+A model.  As in the case of A+A reaction, the value for
$d/4$ exponent obtained here is not correct.  The computer simulation
and $\epsilon$-expansion analysis of this reaction suggest $d/2$
exponent~\cite{KJ1,KJ2}. To see what happens when ${\cal O}(n^3)$
terms are kept in (\ref{GammaAA})-(\ref{GammaAB}) we proceed with
WBGA-II calculation.

\subsection{The WBGA-II approximation}
\label{sec:ABBA+WBGAII}

In the WBGA-II approximation, when all terms are kept in
Eqs.~(\ref{GammaAA})-(\ref{GammaAB}), it is useful to rewrite these
equations in the vector form
\begin{equation}
  \left( \frac{\partial}{\partial t}+2Dk^2 \right)
     \left( 
        \begin{array}{c} 
          \Gamma^{AA}_k \\ [5pt]
          \Gamma^{BB}_k \\ [5pt]
          \Gamma^{AB}_k
        \end{array}
      \right)
      = - {\bf P}(t) 
     \left( 
        \begin{array}{c} 
          \Gamma^{AA}_k \\ [5pt]
          \Gamma^{BB}_k \\ [5pt]
          \Gamma^{AB}_k
        \end{array}
      \right) 
      -
      \left(
      \begin{array}{c}
        \lambda \varphi_{AA} \\ [5pt]
        \lambda \varphi_{BB} \\ [5pt]
        \delta  \varphi_{AB} 
      \end{array}
      \right)
  \label{dtvecGam}
\end{equation}
where the matrix $\bf P$ is given by
\begin{equation}
 {\bf P} = 
    n_A
    \left(
       \begin{array}{ccc}
          4\lambda + 2 \delta u   & 0  & 2\delta  \\
          0  & 4\lambda u + 2\delta &  2\delta u \\
          \delta u    & \delta     &  (2\lambda+\delta)(1+u)
       \end{array}
    \right)
  \label{P}
\end{equation}
with $u=n_B/n_A$.

Vector equation (\ref{dtvecGam}) is very hard to solve
analytically. However, there are some guidelines how to extract late
time asymptotics. At the WBGA-I level it appears that ABBA model and
the A+A model are very similar. In the following it will be assumed
that such similarity can be extrapolated to the presently studied
WBGA-II level.  This implies that mean field behavior should be
expected from Eq.~(\ref{dtvecGam}).

To get the feeling for what follows it is useful to analyze
Eqs.~(\ref{dtnA}) and (\ref{dtnB}) at the mean field level by
neglecting fluctuations and setting $\Phi_{\rho\nu}$ to zero with
$\rho,\nu=A,B$. Carrying out such procedure gives
\begin{eqnarray}
  & & \frac{\partial}{\partial t} n_A = 
        - \left( 
             \lambda n_A^2 + \delta n_A n_B 
          \right) 
          \label{dtnA1} \\
  & & \frac{\partial}{\partial t} n_B = 
        - \left( 
             \lambda n_B^2 + \delta n_A n_B 
          \right)
          \label{dtnB1} 
\end{eqnarray}  
Equations above can be solved approximatively for large $t$ (please see
refs.~\cite{KJ2} for details) and one obtains
\begin{eqnarray}
  & & n_A\sim \frac{1}{\lambda t} 
      \label{mfa} \\
  & & n_B\sim \frac{
                 n_{0,B} 
                   }{ 
                 [ n_{0,A}\lambda t ]^{\gamma} 
               }  
      \label{mfb}
\end{eqnarray}
provided $\gamma\equiv\delta/\lambda>1$ and $n_{0,A}>n_{0,B}$. For
$\delta=\lambda$ ($\gamma=1$) or $n_{0,A}=n_{0,B}$ solution is
trivial, and it can be easily shown that in such case the ABBA model
belongs to the A+A universality class. These simple cases are not
considered here.  Initial imbalance in particle concentration leads to
faster diminishing of minority species, i.e.  $u(t)=n_B(t)/n_A(t)
\rightarrow 0$ as $t\rightarrow\infty$ given $0<u(0)<1$.

In the following we assume the mean field ansatz
(\ref{mfa})-(\ref{mfb}) and try to solve (\ref{dtvecGam}) with it. The
validity of such mean field ansatz will be checked self consistently
at the end.  For large times, and with mean field behavior
($u\rightarrow 0$), the matrix ${\bf P}$ can be approximated by
\begin{equation}
  {\bf P}\approx \lambda n_A {\bf\Pi}
\end{equation}
with
\begin{equation}
  {\bf\Pi} =
    \left(
       \begin{array}{ccc}
          4 & 0       &  2\gamma         \\
          0 & 2\gamma &  0               \\
          0 & \gamma  &  2+\gamma
       \end{array}
    \right)
  \label{Pi}
\end{equation}
The fact that ${\bf P}$ (in the approximate form) is constant matrix
multiplied by time dependent function implies that $[\dot{\bf
P}(t),\dot {\bf P}(t')]=0$ (dot over symbol ${\bf P}$ denotes time
derivative). This being the case, Eq.~(\ref{dtvecGam}) can be treated
as scalar equation and calculation similar to the one in the
subsection \ref{sec:AA+WBGAII} gives
\begin{eqnarray}
  \left( 
    \begin{array}{c}
      \varphi_{AA}(t) \\ [5pt]
      \varphi_{BB}(t) \\ [5pt]
      \varphi_{AB}(t) 
    \end{array}
  \right)
  & = &
  \left( 
    \begin{array}{c}
      n_A(t)^2 \\ [5pt]
      n_B(t)^2 \\ [5pt]
      n_A(t) n_B(t) 
    \end{array}
  \right) - \nonumber \\
 & & 
  -
  \int_0^t dt' {\bf J}(t,t')
  \left( 
    \begin{array}{c}
      \lambda \varphi_{AA}(t') \\ [5pt]
      \lambda \varphi_{BB}(t') \\ [5pt]
      \delta \varphi_{AB}(t') 
    \end{array}
  \right)
  \label{varphi2}
\end{eqnarray}
where matrix ${\bf J}(t,t')$ is given by
\begin{equation}
  {\bf J}(t,t') = G(t,t') {\rm exp} 
    \left[ -\xi(t,t') {\bf\Pi} \right]
  \label{Jtt}
\end{equation}
and 
\begin{equation}
  \xi(t,t')\equiv \lambda \int_{t'}^t dt'' n_A(t'')
  \label{xi}
\end{equation}
Please compare Eqs.~(\ref{varphi1})-(\ref{Itt}) and
(\ref{varphi2})-(\ref{Jtt}). They are very similar, the only difference
being in the matrix character of (\ref{varphi2}-\ref{Jtt}). Following
same steps as in section \ref{sec:AA+WBGAII} the Eq.~(\ref{varphi2})
can be approximated as
\begin{equation}
  \left( 
    \begin{array}{c}
      \varphi_{AA} \\ [5pt]
      \varphi_{BB} \\ [5pt]
      \varphi_{AB} 
    \end{array}
  \right)
   \approx 
  \left( 
    \begin{array}{c}
      n_A^2 \\ [5pt]
      n_B^2 \\ [5pt]
      n_A n_B 
    \end{array}
  \right) 
  -
  {\bf J}(t,0) 
  \left( 
    \begin{array}{c}
      {\cal I}_{AA} \\ [5pt]
      {\cal I}_{BB} \\ [5pt]
      {\cal I}_{AB} 
    \end{array}
  \right)
  \label{varphi3}
\end{equation}
Now we proceed to show that, as in the case of (\ref{varphiapprox}),
the second term on the right hand side of equation (\ref{varphi3}) can
be neglected.  

Matrix ${\bf\Pi}$ can be diagonalized as ${\bf\Pi U}={\bf U \Omega}$. The
${\bf\Omega}$ is diagonal matrix containing eigenvalues
\begin{equation}
  \omega_1=4 \ , \ \ 
  \omega_2=2\gamma \ , \ \ 
  \omega_3=2+\gamma
  \label{eigenvalues}
\end{equation}
and matrix ${\bf U}$ contains eigenvectors
\begin{equation}
  {\bf U} = 
    \left(
      \begin{array}{ccc}
        1 & \frac{\gamma}{\gamma-2} & \frac{2\gamma}{\gamma-2} \\ 
        0 & \frac{\gamma-2}{\gamma} & 0                        \\
        0 & 1                       & 1
      \end{array}
    \right)
  \label{eigenvectors}
\end{equation}
Inserting (\ref{mfa}) into (\ref{xi}), and assuming large $t$, leads to 
\begin{equation}
  \xi(t,0) \sim const + \ln t
  \label{xi1}
\end{equation}
and using (\ref{xi1}) in (\ref{Jtt}) gives
\begin{equation}
  {\bf J}(t,0)\sim {\rm const}\ t^{-d/2} {\bf U} 
                \left(
                  \begin{array}{ccc}
                     t^{-4} &     0        & 0               \\
                     0      & t^{-2\gamma} & 0               \\
                     0      &     0        & t^{-(2+\gamma)} 
                  \end{array}
                \right)
                {\bf U}^{-1}
  \label{Jt0}
\end{equation}
Finally, the second term on the right hand side of (\ref{varphi3}) can
be calculated explicitly. Inserting (\ref{Jt0}) into (\ref{varphi3}),
and assuming that ${\cal I}_{\rho\nu}$ $\rho,\nu=A,B$ are constants
(can be checked for self consistency at the end), results in
\begin{eqnarray}
 & &  \varphi_{AA} \approx n_A^2 
            + t^{-d/2}( c_1 t^{-\omega_1} 
                      + c_2 t^{-\omega_2} 
                      + c_3 t^{-\omega_3} ) \label{c1-3} \\ 
 & &  \varphi_{BB} \approx n_B^2 
            + t^{-d/2} c_4 t^{-\omega_2}  \label{c4} \\ 
 & &  \varphi_{AB} \approx n_A n_B 
            + t^{-d/2}( c_5 t^{-\omega_2} 
                      + c_6 t^{-\omega_3} )  \label{c5-6}
\end{eqnarray}
The explicit form of constants $c_1$, $c_2$, $c_3$, $c_4$, $c_5$ and
$c_6$ is not interesting since aim is to show that terms containing
these constants are sub-leading to the mean field terms.  By studying
equation above row by row, it is possible to show that for $\gamma\ge
1$ terms involving constants are sub-leading to the mean field terms.

To see that terms originating from ${\bf J}(t,0)$ are sub-leading one
really has to calculate ${\bf U}$ explicitly. For example, not knowing that
contribution from $\omega_1$ is absent in (\ref{c4}), there would be a
need to compare $t^{-2\gamma}$ (asymptotics of the mean field $n_B^2$
term in \ref{c4}) with $t^{-(d/2+4)}$ (coming from ${\bf J}(t,0)$ and
$\omega_1$ eigenvalue). One would conclude that
$\gamma=\delta/\lambda$ can not be too large if mean field asymptotics
is to hold. In reality, there is no such bound on ratio
$\delta/\lambda$ since eigenvalue $\omega_1$ does not appear in
Eq.~(\ref{c4}), but this can only be seen after explicit calculation.

Main findings so far is that WBGA describes ABBA and A+A model in the
same way. For both models the WBGA-I (WBGA-II) predict $d/4$ (mean
field) density decay exponents.  In the next section attempt will be
made to improve WBGA method in order to obtain correct value of
density decay exponent for ABBA model.

\subsection{The hybrid of WBGA/Kirkwood applied to A+A and B+B sectors}
\label{sec:ABBA+hybrid}

How deep the weakness of WBGA goes? What needs to be changed in
equations of motion (\ref{GammaAA})-(\ref{GammaAB}) in order to get
correct decay exponent?  To answer these questions we begin by
modifying the A+A and B+B reaction sectors in
(\ref{GammaAA})-(\ref{GammaAB}) by using the recipe from section
\ref{sec:AA+hybrid}. It was already remarked in the section
\ref{sec:model} that contributions to $H$ describing different
reaction sectors enter additively, and this feature is reflected in
equations of motion (\ref{GammaAA})-(\ref{GammaAB}). Because of this
it is possible to focus on terms describing influence of A+A and B+B
reactions, i.e. terms proportional to $\lambda$, which govern decay
exponents of the ABBA model.  At the moment terms proportional to $\delta$
in (\ref{GammaAA})-(\ref{GammaAB}) that describe A+B reaction will not
be changed.

If one is to follow procedure described in section
\ref{sec:AA+hybrid}, the ${\cal O}(n^3)$ term in Eq.~(\ref{GammaAA})
has to be modified as
\begin{equation}
  4 \lambda n_A \Gamma_k^{AA} \rightarrow 
  2\lambda \Gamma_k^{AA} \frac{n_A^2+\Phi_{AA}}{n_A}
\end{equation}
and likewise for Eq.~(\ref{GammaBB})
\begin{equation}
  4 \lambda n_B \Gamma_k^{BB} \rightarrow 
  2\lambda \Gamma_k^{BB} \frac{n_B^2+\Phi_{BB}}{n_B}
\end{equation}
This gives a new set of, hopefully better, equations:
\begin{eqnarray}
  \frac{\partial}{\partial t} \Gamma_k^{AA} 
      & = & - 2 D k^2 \Gamma_k^{AA}   
            - \lambda \left(  n_A^2 + \Phi_{AA} \right) - \nonumber \\
      &   & - 2 \lambda \Gamma_k^{AA} \frac{n_A^2+\Phi_{AA}}{n_A} -\nonumber \\
      &   & - 2 \delta ( n_B \Gamma^{AA}_k + n_A \Gamma^{AB}_k )
      \label{GammaAA1} \\
  \frac{\partial}{\partial t} \Gamma_k^{BB} 
      & = & - 2 D k^2 \Gamma_k^{BB}   
            - \lambda \left(  n_B^2 + \Phi_{BB} \right) - \nonumber \\
      &   & - 2\lambda \Gamma_k^{BB} \frac{n_B^2+\Phi_{BB}}{n_B} - \nonumber \\
      &   & - 2 \delta ( n_A \Gamma^{BB}_k + n_B \Gamma^{AB}_k ) 
      \label{GammaBB1} 
\end{eqnarray}
Equation~(\ref{GammaAB}) stays the same, although the
Eq.~(\ref{GammaAB}) contains term proportional to $\lambda$ which
should be modified if one follows the principle outlined above.
However, at the moment, Eq.~(\ref{GammaAB}) will not be changed.

It is useful to employ similar notation to the one used in section
\ref{sec:AA+WBGAII};
\begin{equation}
  \Gamma_k^{\rho\nu} \equiv n_\rho n_\nu \chi_k^{\rho\nu} 
  \label{chikab}  
\end{equation}
and
\begin{equation}
  \chi_{\rho\nu} \equiv 1 + \frac{1}{V} \sum_{k\ne 0} \chi_k^{\rho\nu} 
  \label{chiab}
\end{equation}
with $\rho,\nu=A,B$.  Using (\ref{chikab}) and (\ref{chiab})
in Eqs.~(\ref{dadt}) and (\ref{dbdt}) results in
\begin{eqnarray}
 & & \frac{\partial}{\partial t} n_A 
       = - ( \lambda n_A^2 \chi_{AA} + \delta n_A n_B \chi_{AB} ) 
     \label{dtnA2}\\
 & & \frac{\partial}{\partial t} n_B 
       = - ( \lambda n_B^2 \chi_{BB} + \delta n_A n_B \chi_{AB} )
     \label{dtnB2}
\end{eqnarray}
Implementing same notation in Eqs.~(\ref{GammaAA1}), (\ref{GammaBB1})
and (\ref{GammaAB}) gives
\begin{eqnarray}
  \frac{\partial}{\partial t} \chi_k^{AA} 
      & = & - 2 D k^2 \chi_k^{AA} - \lambda \chi_{AA} - \nonumber \\
      &   & - 2 \delta n_B ( 1 - \chi_{AB} ) \chi_k^{AA} 
            - 2 \delta n_B \chi_k^{AB}
      \label{chiAA} \\
  \frac{\partial}{\partial t} \chi_k^{BB} 
      & = & - 2 D k^2 \chi_k^{BB} - \lambda \chi_{BB} - \nonumber \\
      &   & - 2 \delta n_A ( 1 - \chi_{AB} ) \chi_k^{BB} 
            - 2 \delta n_A \chi_k^{AB}
      \label{chiBB} \\
  \frac{\partial}{\partial t} \chi_k^{AB} 
      & = & - 2 D k^2 \chi_k^{AB} - \delta \chi_{AB} - \nonumber \\
      &   & - [ 
                \lambda n_A ( 2 - \chi_{AA} ) +
                \lambda n_B ( 2 - \chi_{BB} ) + \nonumber \\
      &   & \ \ \ 
                + \delta ( n_A + n_B ) ( 1 - \chi_{AB} ) 
              ] \chi_k^{AB} - \nonumber \\ 
      &   & - \delta ( n_A \chi_k^{AA} + n_B \chi_k^{BB} )
      \label{chiAB} 
\end{eqnarray}

The numerical solution of the set of equations above is shown in
Fig. 2 (dotted line). The full line is a result of Monte Carlo
simulation where particle densities are obtained as ensemble averages
over 500 runs (simulation is repeated 500 times with a shift in the
random number generator).  The Eqs.~(\ref{chiAA})-(\ref{chiAB}) do not
describe ABBA model properly, not even qualitatively, since minority
species die out faster (the particle density ratio grows to infinity)
while the simulation shows that density ratio should saturate to a
constant value (full line). Thus the attempt of modifying terms
describing A+A and B+B reaction sector in AA and BB correlation
functions by using shortened Kirkwood superposition approximation does
not cure weaknesses of WBGA approach. The inspection of individual
density decays reveals that the equations above correctly describe
decay of majority species, i.e. $n_A\sim ({\rm const}) t^{-d/2}$, but
fail do describe decay of minority species $n_B$.

In the following more terms will be modified by using Kirkwood
superposition approximation.  The dash-dot line in Fig.~2 shows a
solution of equations (not shown here) obtained from modifying A+A and
B+B reaction sectors using shortened Kirkwood superposition
approximation in the Eq.~(\ref{GammaAB}) describing time evolution of
AB correlation function. Equations obtained in this way are identical
to the (\ref{chiAA})-(\ref{chiAB}) with only difference that
Eq.~(\ref{chiAB}) changes in a way that terms proportional to
$\lambda$ drop out. In Fig.~2 it can be seen that even if all
$\lambda$ proportional terms are modified the density ratio curve
climbs to infinity, which indicates that minority species dies out,
which is not correct behavior. However, the overall trend gets better
as the dash-dot curve lies bellow dotted one and is pushed towards
simulation curve.

To continue this line of incremental changes the equations of motion
where studied where even the A+B reaction sector was modified using
shortened Kirkwood superposition approximation (all terms proportional
to $\delta$ were modified). Equations obtained in this way are same as
in (\ref{chiAA})-(\ref{chiAB}) the only difference being in the fact
that all seemingly ${\cal O}(n)$ terms drop out. Thus in the
(\ref{chiAA})-(\ref{chiAB}) only diffusion term and terms $\lambda
\chi_{AA}$, $\lambda \chi_{BB}$ and $\delta \chi_{AB}$ are kept in
(\ref{chiAA}), (\ref{chiBB}) and (\ref{chiAB}) respectively. These
equations are not shown explicitly to save the space but it should be
clear how they look like.  The numerical solution for this set of
equations is shown in Fig.~2 as dashed line. [It is possible to
analytically extract density decay asymptotics for this truncated set
of equations which gives $n_A(t)\sim({\rm const})t^{-d/2}$ and
$n_B(t)\sim({\rm const'})t^{-d/4}$.]

The set of equations where Kirkwood superposition approximation has
been applied fully agrees with the numerical simulation much better
than the rest which are mixture of WBGA and Kirkwood superposition
approximation. This is strong indication that, at least for the ABBA
model, the Kirkwood superposition approximation is superior to the
WBGA method.  For example, in Fig. 2, the trend in all curves improve
as the content of superposition approximation is increased in
equations of motion (dotted line is worse as it climbs fastest,
dash-dot line is a bit better, while only dashed line where
superposition approximation is implemented fully saturates to
constant). As the goal of the present study is to understand WBGA
method better, the more detailed analysis of ABBA reaction based on
Kirkwood superposition approximation will be presented in forthcoming
publication.

\section{Conclusions}
\label{sec:conclusions}

The goal of this study was to test workings of WBGA and to relate
some seemingly independent calculation schemes available in the
literature. In particular, this work has addressed few important
issues.

(1) It was shown that WBGA-I is equivalent to the dressed tree calculation
introduced in ref.~\cite{Lee}, and this equivalence holds for any
model where particles annihilate in pairs.  Furthermore, it was shown
that for the A+A reaction the dressed tree calculation results in
$d/4$ density decay exponent.  Thus the contradictory claims of
refs.~\cite{Lee} and~\cite{rev6} have been sorted out.

(2) The WBGA method describes A+B reaction well, but does not work for
A+A and ABBA models, and it is reasonable to expect that there are
more models that can not be described by WBGA.  In the case of ABBA
model WBGA predicts faster vanishing of minority species which is
suggestive of the A+B like behavior rather than the behavior of the
ABBA model as found in refs.~\cite{KJ1,KJ2}. This bias towards A+B
type behavior is very hard to get rid off as successively correcting
more and more terms in equations of motion for ABBA model by using
Kirkwood superposition approximation results in faster vanishing of
minority species. The vanishing of minority species persists until all
terms are modified by Kirkwood superposition approximation. The WBGA
emphasize A+B reaction sector too strongly in the ABBA model.

(3) Findings of this work suggest that formalism employed by
Mattis-Glasser in~\cite{rev6} where small $n_0$ expansion is
introduced (and applied to study A+B model) is somewhat
questionable. This procedure works on A+B model, but might not work
for other models. It can be shown (by rescaling $a^\dagger
n_0\rightarrow a^\dagger$ and $a/n_0\rightarrow a$) that for the type
of models studied in here, the small $n_0$ approximation of
ref.~\cite{rev6} amounts to taking away three body terms in
Hamiltonian given in Eq.~(\ref{Hreactx'}) or (\ref{Hreactk})
(e.g. operators of the type $a^\dagger a a$ and likewise any mixture
of $a$ or $b$ operators). The neglect of these terms amounts exactly
to WBGA-I approach, i.e. neglect of seemingly ${\cal O}(n^3)$ terms in
equations of motion for correlation functions; for example, equations
(\ref{dn3})-(\ref{varphi}) of section \ref{sec:AA+WBGAI} (A+A), or
equations (\ref{dadt})-(\ref{dvarab}) from section
\ref{sec:ABBA+WBGAI} (ABBA). In here it has been shown that WBGA-I set
of equations result in incorrect $d/4$ density decay exponent when
used for A+A and ABBA models. For these reasons, small $n_0$
expansion, which effectively means taking away three body term in
Hamiltonian, can not be trusted if used beyond A+B model.

(4) The Kirkwood superposition approximation was formulated for the
case of lattice dynamics with multiple occupancy of lattice sites
allowed. There is strong indication from the present analysis that
Kirkwood superposition approximation is superior to the WBGA method,
at least when applied to the A+A and ABBA models.  However, one has to
be careful in claiming superiority of Kirkwood superposition
approximation over WBGA approach since in here it was shown that
Kirkwood superposition approximation describes ABBA model
qualitatively, but it remains to be seen whether it works on the A+B
model which WBGA describes well. Similar study (where only single
occupancy of lattice sites was allowed and reaction range was assumed
short but finite) showed that Kirkwood superposition approximation can
describe A+B reaction~\cite{rev4}, and claim of superposition
approximation superiority is likely to be correct but, nevertheless,
such claim has to be tested throughly.

The application of Kirkwood superposition approximation to the ABBA
and A+B models will be presented in the forthcoming publication as
there are many interesting technical details that need to be sorted
out. For example, from present study there are some indications
(results not shown here) that when using Kirkwood superposition
approximation for extremely short range reaction, such as on-site
reaction studied here, care has to be taken so that thermodynamic
limit is accounted for properly. It seems that one has to approach
Kirkwood superposition approximation through formalism developed in
section \ref{sec:AA+hybrid}.

To conclude, it would be interesting to have a relatively simple
approximation at hand, not far away from pair approximation, which
could be used to extract qualitative asymptotics for arbitrary
reaction-diffusion model. Clearly such program is ambitious since in
reality one is bound to make approximation which is related to the
particular model but, nevertheless, it is worth a try. The A+A and A+B
reaction-diffusion models (or combination of them) are excellent
bench-mark models and any successful approximation should strive to
describe these reactions properly.  Present study is a attempt in this
direction.

\begin{acknowledgments}

I would like to thank Prof. Malte Henkel and the staff at the
Laboratoire de Physique des Materiaux, Universite Henri Poincare Nancy
I, for warm hospitality where part of this work was done.

\end{acknowledgments}

\appendix

\section{Approximation for $G(\lowercase{t}-\lowercase{t}')$}
\label{app:Gtt}

In Eq.~(\ref{Gtt}) sum over $k$ can be approximated as integral
\begin{equation}
  G(t-t') = \frac{1}{(2\pi)^d} 
    \left[ 
       \int_{-\pi}^\pi dp e^{-2Dp^2(t-t')}
    \right]^d
  \label{Gtt1}
\end{equation}
The equality sign is strictly valid in thermodynamic limit only when
$V\rightarrow\infty$ has been taken.  It is useful to approximate
integral above further by changing bounds of integration
\begin{equation}
  G(t-t') \approx \frac{1}{(2\pi)^d} 
     \left[
       \int_{-\infty}^\infty dp
       e^{-p^2 \Lambda^{-2}-2D\kappa^2(t-t')}
     \right]^d
  \label{Gtt2}
\end{equation}
$\Lambda$ corresponds to a large cutoff for the $k$-vector summation
(it regularizes UV divergences of theory). The comparison of
(\ref{Gtt1}) and (\ref{Gtt2}) quickly reveal that $\Lambda$ is a
constant roughly equal to $\pi$. The evaluation of integral above gives
Eq.~(\ref{Gtt3}) with $\eta=\frac{1}{2D\Lambda^2}$. It is convenient
to chose $\eta=\frac{1}{8\pi D}$ and $\Lambda=2\sqrt{\pi}$ since this
gives correct value for $G(0)=1$. For large $t-t'$ particular value of
$\Lambda$ appears to be irrelevant, {\em i.e.} absent from final
expressions for particle density for large times. However, this really
depends on the critical dimension of the field theory.

\section{Integration of singular kernel}
\label{app:kernal}

In here the numerical treatment of Eqs.~(\ref{dn3}-\ref{varphi}) is
described in a more detail. The general procedure for integrating
expressions of the type
\begin{equation}
  I_i[f] = \int_0^{t_i} ds K(t_i,s)f(s)
  \label{Iif}
\end{equation}
where $K(t,s)$ is singular when s approaches t is described in
ref.~\cite{delves}. The $i=0,1,2,\ldots$ and $t_i = i h$. In here the
particular focus in on the particular form of singular kernel
originating from propagator $G(t,t')$;
\begin{equation}
  K(t,s)=(t-s+\eta)^{-\alpha}
  \label{Kts}
\end{equation}
The aim is to find quadrature formula which can integrate (\ref{Iif})
exactly if $f(s)$ is linear within intervals $t_{i-1},t_i$. Thus goal is
to find coefficients $w_{ij}$ in
\begin{equation}
  I_i[f] \approx \sum_{j=0}^i w_{ij} f(t_j)
  \label{Iifapp}
\end{equation}
such that above equation turns equality for piecewise linear function
$f(s)$. Implementing this procedure gives
\begin{equation}
  I_i[f] = \sum_{j=0}^{i-1} \int_{s_j}^{s_{j+1}} ds K(t_i,s) f(s)
\end{equation}
and after approximating $f(s)$ as linear function in intervals
$t_j,t_{j+1}$ for $j=0,..,i-1$
\begin{equation}
  f(s) \approx \frac{s_{j+1}-s}{h} f(s_j) + \frac{s-s_j}{h} f(s_{j+1})
\end{equation}
gives (\ref{Iifapp}) with
\begin{eqnarray}
  w_{i0} &=& \int_0^h ds K(t_i,s) \frac{h-s}{h} \\
  w_{ij} &=& \int_{s_{j-1}}^{s_j} ds K(t_i,s) \frac{s-s_{j-1}}{h}
                 + \nonumber \\ 
         & &     + \int_{s_j}^{s_{j+1}} ds K(t_i,s) \frac{s_{j+1}-s}{h} \\
  w_{ii} &=& \int_{s_{i-1}}^{s_i} ds K(t_i,s) \frac{s-s_{i-1}}{h}
\end{eqnarray}
and after performing integrals above with $K(t,s)$ given in (\ref{Kts})
\begin{eqnarray}
  w_{i0} &=& \frac{h^{1-\alpha}}{(\alpha-1)(\alpha-2)}
             [(2-\alpha-i-\eta)(i+\eta)^{1-\alpha} + \nonumber \\
         & &  + (i+\eta-1)^{2-\alpha}] 
             \label{wi0} \\
  w_{ij} &=& \frac{h^{1-\alpha}}{(\alpha-1)(\alpha-2)}
             [(i-j+\eta-1)^{2-\alpha} + \nonumber \\
         & & + (i-j+\eta+1)^{2-\alpha} - 2(i-j+\eta)^{2-\alpha}] 
             \label{wij} \\
  w_{ii} &=& \frac{h^{1-\alpha}}{(\alpha-1)(\alpha-2)}
             [(1+\eta)^{2-\alpha}+ \nonumber \\
         & & + \eta^{1-\alpha}(\alpha-\eta-2)]
             \label{wii}
\end{eqnarray}
The pair of equations in (\ref{dn3}-\ref{varphi}) is discretized as
follows. First the differential equation (\ref{dn3}) is rewritten in
integral form as $n(t)=n_0-\lambda \int_0^t ds \varphi(s)$ and
trapezoidal rule is used to evaluate integral since all functions are
well behaved. However, for Eq.~(\ref{varphi}) the rule (\ref{Iifapp})
and (\ref{wi0})-(\ref{wii}) designed for singular kernel is
used. Implementation of this philosophy gives
\begin{eqnarray}
  n_i = n_0 - \lambda h [ \frac{\varphi_0}{2} 
              + \sum_{j=1}^{i-1} \varphi_j + \frac{\varphi_i}{2} ] \\
  \varphi_i = n_i^2 - \lambda \sum_{j=0}^{i-1} w_{ij} \varphi_j 
                    - \lambda w_{ii} \varphi_i 
\end{eqnarray}
where $n_i=n(t_i)$ and $\varphi_i=\varphi(t_i)$ for
$i=0,1,2,\ldots$. Given that all $n_j$ and $\varphi_j$ are known for
$j=0,1,2,\ldots,i-1$ using equations above it is possible to calculate
$n_i$ and $\varphi_i$. The iteration is started with $n_0=n(0)$ and
$\varphi_0=n_0^2$.

\section{Finding inverse Laplace transform of $\lowercase{\kappa(s)}$ 
for $\lowercase{d}=2$}
\label{app:kappa}

In here inverse Laplace transform of $\kappa(s)$ for $d=2$ given in
Eq.~(\ref{kappas}) will be found. Due to the presence of log one has
to use Bramowitz contour to perform integration over $s$. Also,
function $\kappa(s)$ does not have poles. This means that only
contribution to $\kappa(t)$ comes from the branch cut and one obtains
\begin{equation}
  \kappa(t) = 8\pi D \int_0^\infty \frac{du}{u} e^{-ut}
      \frac{1}{(\gamma_E+\ln(\eta u))^2+\pi^2}
\end{equation}
It is useful to re-scale integration variable as $\eta u\rightarrow u$
with only change that time variable appears in combination $\tau\equiv
t/\eta$. In the following we set $\eta=1$ but keep in mind that at the
end of calculation $t$ has to be changed into $t/\eta$.

As $t$ grows, due to the presence of ${\rm exp}(-ut)$, only smaller and
smaller values for $u$ contribute to the integral above and it is
useful to split integration over $u$ in the two intervals (a) from $0$
to $c$ and (b) from $c$ to infinity, where $0<c<1$;
$\kappa(t)=\kappa_a(t)+\kappa_b(t)$. It is possible to find upper
bound for $\kappa_b$ as follows. Starting from
\begin{equation}
  \kappa_b(t) \le \frac{8\pi D}{\gamma_E^2+\pi^2} 
                 \int_c^\infty \frac{du}{u} e^{-ut}    
\end{equation}
which, upon using the fact that for particular range of integration
above $1/u\le 1/c$ and performing remaining integration, gives
\begin{equation}
  \kappa_b(t) \le \frac{8\pi D}{(\gamma_E^2+\pi^2)c} \frac{e^{-ct}}{t} 
\end{equation}
It will be shown that $\kappa_a(t)$ vanishes lot slower than $\kappa_b(t)$. 
One can approximate expression for $\kappa_a$ as 
\begin{equation}
  \kappa_a \approx (8\pi D) \int_0^c \frac{du}{u} e^{-ut} \frac{1}{(\ln u)^2}
\end{equation}
and it can be shown that terms omitted do not influence leading order
behavior for $\kappa_a$. By using partial integration expression above becomes
\begin{equation}
  \kappa_a = 8\pi D \left[ 
     - t \int_0^c du e^{-ut}\frac{1}{\ln u} + {\cal O}(e^{-ct}) \right]
\end{equation}
The integral over $u$ is most conveniently calculated by changing
variables $tu=v$ which gives
\begin{equation}
  \kappa_a(t) \approx \frac{8\pi D}{\ln t} \int_0^{ct} dv e^{-v} 
                      \frac{1}{1-\frac{\ln v}{\ln t}}   
\end{equation}
By sending upper integration limit to infinity, and expanding
denominator in series over $\ln v/\ln t$ one gets
\begin{equation}
  \kappa_a = \frac{8\pi D}{\ln t} [ 1 + {\cal O} (1/\ln t) ] 
\end{equation}
Keeping in mind that transformation $t\rightarrow t/\eta$ has to be
made in the expression above gives result for $\kappa(t)$ in
Eq.~(\ref{kappat}).

\begin{figure}
\epsfxsize=9cm
\epsfysize=8cm
\epsfbox{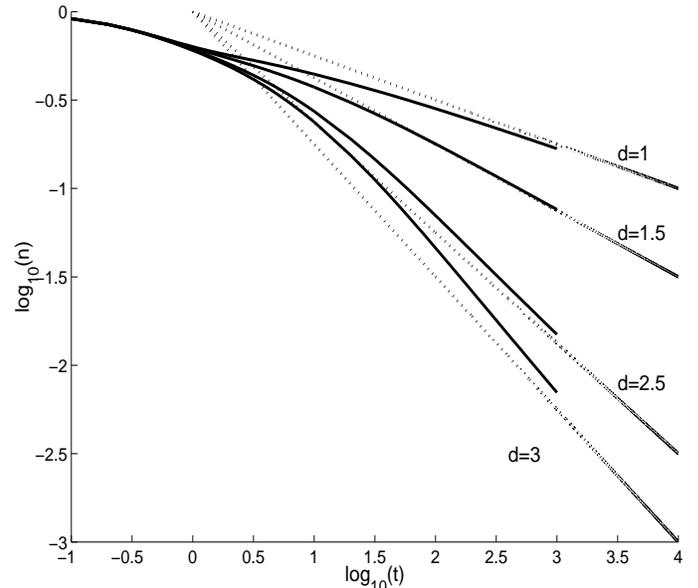}
\caption{FIG 1. The numerical solution of
Eqs.~(\ref{dn3}-\ref{varphi}) for $d=1,1.5,2.5,3$ (solid lines). The
dotted lines indicate asymptotics as given by Eq.~(\ref{ntasym}). Time
is given in seconds and particle density $n(t)$ is dimensionless in
units of particles per site. Initial density $n_0$ was set equal to 1,
and reaction rate $\lambda=1s^{-1}$ was used.  }
\label{fig1}
\end{figure}

\begin{figure}
\epsfxsize=9cm
\epsfysize=8cm
\epsfbox{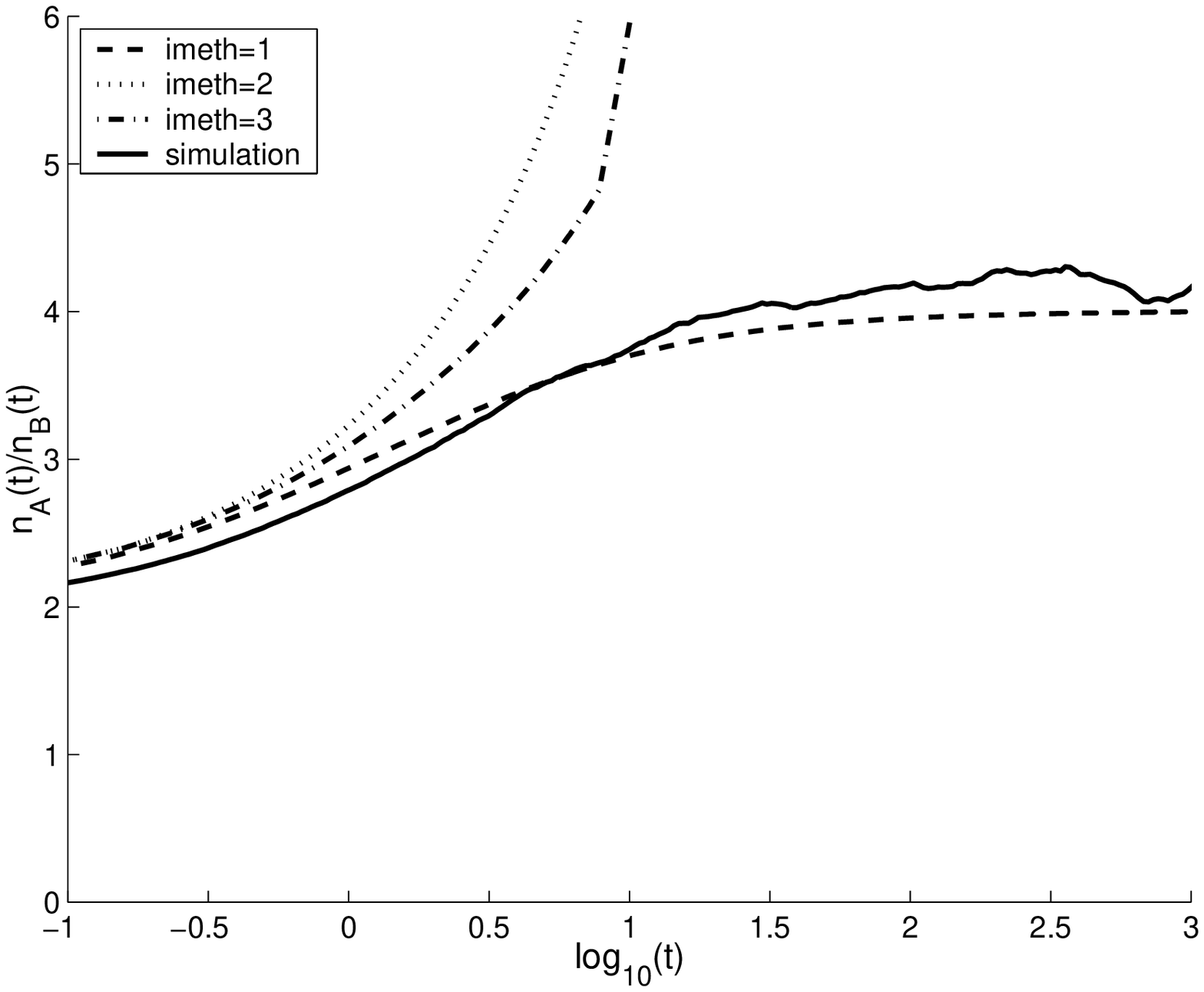}
\caption{FIG 2. The numerical solution of
Eqs.~(\ref{dtnA2}-\ref{chiAB}) for $d=1$ with increasing amount of
Kirkwood superposition approximation embedded (dotted, dash-dot and
dashed line). Full curve is result of a Monte Carlo simulation
(average of 500 runs). Parameters used are $L=1000$, $n_A(0)=2$,
$n_B(0)=1$, $\lambda=1$ and $\delta=2$.  }
\label{fig2}
\end{figure}

\end{multicols}


\begin{thebibliography}{19}
\bibitem{foot2} For example, one way to prepare system initially is to
take given number of particles and distribute them randomly one by one
on the lattice. This way of preparation leads to Poisson distribution
of particle number at each lattice site. Also, it is clear that
preparing system in this way does not lead to correlation among
particles. Thus, saying that particles are distributed according to
Poisson-distribution amounts to saying that there are no correlations
among them.
\bibitem{foot1} The initial state of the system was prepared by
allowing for birth and annihilation of particles and waiting long
enough to establish stationary state. Once this stationary state was
reached, particle birth was ceased and system continued to evolve by
annihilation process solely.
\bibitem{foot3} Acting by time derivative on (\ref{BarPsiHPsi0}) gives,
up to minus sign, $\langle 0|a_0\bar H|\bar\Psi\rangle$ which due to
$\langle 0|\bar H=0$ (probability conservation) can be rewritten as
$\langle 0|a_0\bar H-\bar Ha_0|\bar\Psi\rangle$.
\bibitem{foot4} Inverse Fourier transform of $\langle
a_{k_1}a_{k_2}a_{k_3} \rangle$ is defined as $\langle a_x a_y a_z
\rangle = \frac{1}{V^{3/2}} \sum_{k_1,k_2,k_3}e^{i(k_1x+k_2y+k_3z)}
\langle a_{k_1}a_{k_2}a_{k_3} \rangle$.
\bibitem{foot5} The Fourier transform of (\ref{kirkwood}) is
calculated by using $\langle a_x a_y \rangle = \frac{1}{V} \sum_k
e^{ik(x-y)} \langle a_k a_{-k} \rangle$. Thus, translational
invariance is ensured in form of $\langle a_x a_y \rangle$. The term
$\delta(k_1+k_2+k_3)$ appears in (\ref{kirkwoodk}) as artifact of
that.
\bibitem{foot6} The density decay amplitude for A+A model is
independent of $\lambda$ and $n_0$ while the one of A+B model depends
only on $n_0$ (given that systems are observed bellow critical
dimension, naturally).
\end{thebibliography}
\end{document}